\RequirePackage{fix-cm}
\documentclass[smallextended]{svjour3}       
\smartqed  
\usepackage{draftwatermark}
\SetWatermarkText{DRAFT}
\SetWatermarkScale{1.5}

\usepackage{graphicx}

\usepackage{url}

\usepackage{rotating}

\usepackage{subfig}

\usepackage{multirow}
\usepackage{booktabs}
\newcommand{\tabitem}{~~\llap{\textbullet}~~}
\usepackage{lscape}
\usepackage{rotating}
\usepackage{url}
\usepackage{tikz}
\usepackage{pgfplots}
\usepgfplotslibrary{statistics}
\pgfplotsset{
   boxplot/every box/.style={black, solid,fill=gray},
   boxplot/every whisker/.style={black, solid},
   boxplot/every median/.style={black, solid, very thick},
}
\usepackage{color}
\usepackage{comment}
\usepackage{wasysym}
\usepackage{enumitem}
\usepackage{forloop}
\usepackage{ifthen}
\newcommand{\Qq}[1]{\textbf{#1}}
\newcommand{\QO}{$\Box$}

\newcounter{ql}

\newcounter{qr}
\newcommand{\Qrating}[1]{\QO\forloop{qr}{1}{\value{qr} < #1}{---\QO}}

\newlength{\qt}
\newcommand{\Qtab}[2]{
\setlength{\qt}{\linewidth}
\addtolength{\qt}{-#1}
\hfill\parbox[t]{\qt}{\raggedright #2}
}
\newcounter{itemnummer}
\newcommand{\Qitem}[2][]{
\ifthenelse{\equal{#1}{}}{\stepcounter{itemnummer}}{}
\ifthenelse{\equal{#1}{a}}{\stepcounter{itemnummer}}{}
\begin{enumerate}[topsep=2pt,leftmargin=2.8em]
\item[\textbf{\arabic{itemnummer}#1.}] #2
\end{enumerate}
}
\definecolor{bgodd}{rgb}{0.8,0.8,0.8}
\definecolor{bgeven}{rgb}{0.9,0.9,0.9}
\newcounter{itemoddeven}
\newlength{\gb}
\newcommand{\QItem}[2][]{
\setlength{\gb}{\linewidth}
\addtolength{\gb}{-5.25pt}
\ifthenelse{\equal{\value{itemoddeven}}{0}}{%
\noindent\colorbox{bgeven}{\hskip-3pt\begin{minipage}{\gb}\Qitem[#1]{#2}\end{minipage}}%
\stepcounter{itemoddeven}%
}{%
\noindent\colorbox{bgodd}{\hskip-3pt\begin{minipage}{\gb}\Qitem[#1]{#2}\end{minipage}}%
\setcounter{itemoddeven}{0}%
}
}
\begin{document}

\title{A device-interaction model for users with special needs 
}


\author{Juan Jesus Ojeda-Castelo \and
        Jose A. Piedra-Fernandez \and 
        Luis Iribarne 
}


\institute{
          \and 
          Juan J. Ojeda-Castelo (corresponding author) \and Jose A. Piedra-Fernandez \and Luis Iribarne 
           \at Applied Computing Group, Department of Informatics, University of Almeria, Spain
         }

\date{Received: date / Accepted: date}

\maketitle

\begin{abstract}
Interaction is a fundamental part of using any computer system but it is still an issue for people with special needs. In order to improve this situation, this paper describes a new device-interaction model based on adaptation rules for user models. The aim is the adaptation at the interaction level, taking into account the interaction device features in order to improve the usability through the user experience in the education sector. In the evaluation process, several students from a special education center have participated. These students have either a physical or sensory disability or autism. The results are promising enough to consider that this model will be able to help students with disabilities to interact with a computer system which will inevitably provide tremendous benefits to their academic and personal development.
\end{abstract}

\section{Introduction}
\label{intro}
The European Statistics Department\footnote{European Statistics Department: \url{http://ec.europa.eu/eurostat}} asserts that people with disabilities tend to drop out of education before ending secondary school. In 2011, 1 out of every 4 people were in this situation\footnote{Eurostat Disability Statistics: \url{https://bit.ly/3h3w5mY}}, in particular, the number of school dropouts of disabled people ranged from 11\% in Sweden to more than 60\% in Turkey and Bulgaria.
In order to overcome this situation, it is necessary for the teaching methodology to be inclusive for and adaptive to the user. In this way, it is hoped that students will not find it difficult to study and they will not drop out of school. Natural interaction is a way to solve this problem because many undergraduates with disabilities are unable to use a computer system with the input devices which are currently available. Nowadays, the Internet has become the common means with which to seek answers when the user has any kind of doubt. Natural interaction is very useful in this context since it allows users to control the proper tools by gestures, voice or tactile interaction. There are two assistance methods in this device: the adaptation of educational systems \cite{alquraini2012critical} and the use of serious games for the learning process \cite{ypsilanti2014serious}.

In connection with the adaptation of educational systems the e-learning platforms have a significant role since one of their aspects is that they are adaptable and interactive. One of the ways to adapt hypermedia applications is through user models \cite{de1999adaptive}. Recently, the extent of the smart learning environment based on student models has increased because these contain relevant information for the learning process such as the students prior knowledge or their learning styles \cite{desmarais2012review}. The goal of student models is the customization of the student characteristics with respect to the cognitive, emotional and behavioral variables from the learning procedure. These models can be created in different ways: based on ontologies \cite{labib2017way} to represent the personal and context features or also through the use of serious games \cite{khenissi2015learner}.

In \cite{premlatha2015learning} the authors state that the basis of the adaptation should not only be the learner requirements but also the learning content. In special educational needs, mobile apps have been developed for students with disabilities due to the increasing demands from teachers who are using tablets as a tool for learning in the classroom since these have a myriad of advantages in the educational field \cite{pellerin2013inclusion}. This type of application is adaptable as the content or the user interface can be modified by the tutor \cite{fernandez2013mobile}. In addition, they can be adapted according to the user profile.

In this paper, we explain a system with a user model and a \textit{Device-Interaction Model} that takes into consideration the user interaction and is a mechanism of adaptation together with the rule-based expert system. The aim is to enable students with special needs to interact with a computer system through this adaptable proposal, regardless of their disabilities. The inclusion of game-based mechanics have been helpful for the development of an activity set which allows the pupils to develop cognitive skills and keep them motivated while they try the system.

The main aim is for people with special needs to be able to interact with a computer system regardless of their skills and facilitate intuitive interaction. The specific goals for this project have been:

\begin{itemize}
\item[(a)] Designing an interactive adapted system for people with special needs.
\item[(b)] Creating an intelligent adaptive module that is enabled to choose the most
suitable way of interaction for the user.
\item[(c)] Designing an user model which allows the system to manage the user characteristics in order to customize the system.
\item[(d)] Designing a model focused on the device to improve the user experience.
\end{itemize}

The remainder of the paper is organized as follows: Section \ref{stateArt} presents
some background on Gesture recognition. Section \ref{proposal} describes the proposed
approach. Section \ref{results} shows the results obtained through carrying out an evaluation. Section \ref{conclusions} summarizes the conclusions and discusses future work.

\section{Related Work} \label{stateArt}

Natural interaction in education is related mainly to special educational needs as it is a more convenient way to interact with the system \cite{bartoli2013exploring} than traditional methods. This interaction is a step forward to achieving the goal of a flexible teaching methodology for students with impairments and facilitate inclusion in the classroom. Despite the fact that natural interaction helps the inclusion of students with physical disability, there are cases that focus on the adaptation at the cognitive level for learners who are mentally- challenged. Nevertheless, without a conducive and stimulating environment, efforts to adapt the system will be useless. Therefore, this system is designed as a serious game for the learners to develop their physical abilities and stimulate their cognitive skills while experiencing performing the activities as a game. Some projects which are significantly related to this topic are described below.

\subsection{Natural Interaction in Education} \label{sub_NaturalInteraction}

Natural interaction was applied as a platform for special education with the MS Kinect device. This device encourages interaction and learning through multisensory skills that allow the student to work with kinaesthetic memory. The addition of gesture recognition in the classroom encourages the students to interact with their surroundings. Furthermore, real situations  are  simulated with the gesture recognition, for example, a simulated exercise where the learner has to drive a vehicle and learn the road-safety rules. This methodology assists in retaining their attention and motivating them to learn more.

Kinems \cite{altanis2013children} is a serious game that employs body motion to complete each activity. It is designed for learners with ADHD, autism, dyspraxia and students with learning problems such as dyslexia and dyscalculia. This game requires body and hand motion in order to develop skills such as hand-eye coordination in special education. Users also develop other particular skills when they operate with Kinems such as problem resolution, sequencing, concentration ability and short-term memory. This system provides activities to teach Mathematical concepts and calming activities for stress release and to develop motor skills with a game-based methodology with the purpose of encouraging the students to complete the exercises. Recently, research was conducted in order to study kinetic and learning analytics and show the results to the community \cite{kourakli2017towards}. This study was carried out in two primary schools where 20 students with special needs participated in the experiment. According to the results, Kinems may be an effective solution to improve motor and cognitive skills in students with special needs and a useful tool for inclusive school environments.

In \cite{faisal2016innovating} the motivation of the proposed system is to enable students with special needs to interact in class in spite of their characteristics and also to help in the academic aspect. Due to their differences these students are often excluded by the other students therefore Kinect is used to eliminate these barriers in the educational environment. The device developed by Microsoft was integrated in the system in order to recognize gestures so the students could answer the academic questions using gestures in this simulated environment. The users who participated in the study had Cerebral Palsy (CP), Dyslexia and slow learning. The authors state that the results were remarkable according to the surveys carried out following the experiment and that the system did not present any difficulties to the students after a few interactive sessions

In \cite{hsiao2016using} the authors designed a learning game-based method with gesture recognition for preschool children. The purpose of this method is for the pupils to improve their academic performance and their motor skills, especially their agility and coordination. This system uses the Asus Xtion PRO device to recognize user gestures. The evaluation has two tasks: to assess academic performance and to measure motor skills. In this study, the control group was asked to employ the traditional learning methodology with the students while the experimental group was set up to evaluate the learning process through the game-based method with gesture recognition. The results demonstrated that the experimental group achieved better results than the control group, concluding that this gesture-based learning methodology is more effective than the conventional one.

In \cite{cai2018case} the aim of the study was to see if gesture-based games could improve the learning process in children with autism. Thus, two gesture- based games were created. The Leap Motion device was included to interact with the application. The users had to perform certain gestures depending on the position of  the fingers and the angle between them to achieve the goals in the games. Regarding the results, the students improved their performance with respect to fine motor skills and recognition by using these gesture-based games.

\subsection{Adaptation to Education} \label{sub_Education}

In the field of education, it is important that the teaching method is personalized according to each student’s characteristics and different learning styles. Thus, it is highly recommended that the educational process is  flexible  enough to adapt to the students’ aptitudes. Depending on the learning styles of a student, it may be more suitable to integrate a visual or kinaesthetic learning experience. Hence, the teaching and learning method should be flexible enough to allow the users to develop their skills in a more effective way. The following describes a few studies which include an adaptation mechanism to improve the quality of a learning process.

Topolor is an e-learning flexible system that integrates social interaction \cite{shi2013topolor}. This system is based on a layered architecture, similar to the Dexter model \cite{halasz1994dexter}, which is divided into two basic layers \cite{shi2013social}: the storage layer and the performance layer. The difference between this storage layer and other designs is that it contains a model which facilitates user interaction and it is also compatible with collaborative learning. Furthermore, it has a user model that saves the user preferences and the knowledge model. The performance layer has the objective of analyzing the adaptation strategies to show the learning themes clearly, monitor user activities and update the user models. The system’s adaptability enables it to make changes depending on user interaction \cite{shi2013evaluation}. The system adaptation is focused on learning. For this reason, Topolor has an adaptation paradigm for the learning which it is based on: a learning adaptation plan, adaptive contents about the learning process and peer recommendations about learning.

In \cite{ghiani2015dynamic}, a framework was developed in order to support e-learning adaptive applications with the user’s physiological data. The architecture of this system is composed of an e-learning application, an adaptation motor, user model management, an eye-tracking device and a brain-computer interface (BCI) device. The flexibility of the rule-based expert system is the underlying theme of the dynamic adaptation of the system. The technique applied to the adaptation of this project is the adaptation motor which includes several flexible rules, written in XML, which are based on physiological dynamic data obtained from the user. This system incorporates a multimodal interaction, brain-computer interface and an eye tracker. The brain-computer interface employs NeuroSky\footnote{NeuroSky: \url{http://neurosky.com/biosensors/eeg-sensor/}} which is biosensor enabled to measure brainwaves, concentration ability and attention span \cite{katona2014evaluation}. The eye tracking is applied with the Tobii\footnote{Tobii: \url{http://www.tobii.com/}} device, in which SDK detects the direction the user is looking according to the screen coordinates. The user model is defined in this sequence: 
$$(S, T, N, E, L, A)$$
\noindent which refers to:
\begin{itemize}
\item [S:] Screen position where the user is looking at.
\item [T:] Time the users look outside of the attention area.
\item [N:] Number of eye blinks in three minutes.
\item [E:] Paragraph element being watched by the user.
\item [L:] Time that the user spends looking at a paragraph .
\item [A:] Average attention level.
\end{itemize} 

TECH8 is a smart, flexible and personalized e-learning system \cite{dolenc2015tech8}. Some of its main features are related to the content that guides the user according to their knowledge and learning ability; the content difficulty is adjusted automatically according to  the  user  and  the  structure is  made  in  modules to facilitate new content inclusion. The system includes an intelligent and adaptable part identified as an agent that suggests recommendations for the students to improve their learning process. The evaluation results showed that the use of TECH8 enhanced the student performance by 18\% in comparison to traditional lessons.

UZWEBMAT \cite{ozyurt2013design} is an intelligent, flexible and customized system based on learning styles. This system was designed to teach permutations, combinations and other probability concepts that are taught in mathematical subjects \cite{ozyurt2013integration}. UZWEBMAT is adapted to  the  learning  characteristics  of the students. The learning styles that it considers are visual, hearing and kinaesthetic. To perform this adaptation, UZWEBMAT is created by an expert system that chooses the optimum learning style \cite{ozyurt2012integrating} according to the user response. This is a rule-based expert system where the system procedures are determined by experts in the field in which the content is prepared. To validate that the learning process is more effective with UZWEBMAT, an experimental group using this tool was set up together with a control group using the traditional learning method. The experiments demonstrated that the students from the experimental group were more competent than the ones in the second \cite{ozyurt2014effects}.

In \cite{cinquin2020designing} the process to create a framework whose aim is to make easier the design of accessible e-learning systems were explained. The first step to create this framework was to apply a method called participatory design process where stakeholders are involved in the methodology. In this case, students with schizophrenia, ADHD and dyslexia from six different universities and thirteen experts from different fields related to the topic of this project participated in the procedure. 

The participatory design process had four phases:
\begin{enumerate}
    \item[a)] Requirement Gathering: In this phase, meetings with the professionals and the students with cognitive impairments were organized in order to identify the user needs and preferences.
    \item[b)] Ideas Generation: Co-design sessions were planned to contribute ideas that can be useful for the accessibility in learning environments. These ideas will be really important to implement a prototype in the next phase.
    \item[c)] Prototyping: Once the essential information has been gathered and there are some solutions for the problem outlined, it is necessary to develop a prototype with the aim of checking the ideas that are relevant and which ones should be discarded. At the beginning a paper prototype was created to share it with the rest of the team and afterwards a software prototype was developed.
    \item[d)] Evaluation: In the last phase, the prototype built in the previous step was evaluated to check its validity through a multiple-case study. The participants of this experiment had to fill many questionnaires related to: usability, user experience, self-determination, cognitive load.
\end{enumerate}
  
  After completing this process, the authors were able to build an integrative framework to design accessible learning environments with the feedback obtained from the participatory design process.

\subsection{Serious Games in Education} \label{sub_SeriousGames}

Serious games have been implemented in education systems for primary education, for instance, in Mathematics or Language, to university level with subjects as Commercial Management or Production Efficacy \cite{pourabdollahian2012serious}. They are gaining popularity in the education field since they motivate the students through creating a fun learning environment \cite{de2014serious,dias2019using}. This method is directly related to academic achievements and it has been verified to be beneficial in Languages, History and Physical Education learning \cite{young2012our}. In order to confirm this statement, several studies have also been conducted to assess if serious games can be a useful mechanism for education \cite{backlund2013educational,girard2013serious}, because they have to be designed in a way in which they can transfer the learning to the students while simultaneously keeping them entertained. Furthermore, there is empirical evidence about game-based learning effectiveness \cite{connolly2012systematic}. However, this effectiveness is a result of balancing all the following serious game elements: didactic content, game aspects, game loop, information session, educational values, transference of the learned skills and intrinsic motivation \cite{guillen2012serious}. The content of serious games is usually focused on the topics of the serious games for instance, health aid \cite{afyouni2017therapy}, rehabilitation \cite{hocine2015adaptation}, marketing \cite{blumberg2013serious} or military practice \cite{laamarti2014overview}.

A serious game developed for education is likely to be aimed at the learning of mathematical or linguistic concepts. Although the creation of a specific serious game according to the content of the lesson may be a tedious and complicated task, the community creates tools to solve this issue. For instance, the XR-Serious Games Toolkit \cite{zarraonandia2018toolkit} is a platform whose aim is to reduce the time of implementation and design of cross-reality serious games and it makes the development of these type of applications more suitable for educational purposes. Moreover, these materials need to include more content pertaining to the development of practical skills that will be useful for the user in daily life. From this point of view, this study \cite{brown2011designing} has combined the Serious Game with the services based on the localization for people with intellectual disabilities because they allow the users to acquire skills to travel on their own as well as be confident of learning new paths. These skills are considered to be enabling with regard to achieving independence in life and gaining access to the world of work. Other examples are described below.

 ATHYNOS \cite{avila2018athynos}  is a serious game based on augmented reality whose goal is to improve the motor skills and hand-eye coordination in children with Dyspraxia. In this prototype Kinect and 3D virtual environments were included in order to create a natural interface therefore the students could interact with this interface through gesture recognition. In order to start with ATHYNOS the students had to choose an avatar and then a set of scenarios in which different therapies are shown to the user. These activities had three different levels depending on the user’s skills. In the experiments two groups participated: the control group and the experimentation group. The control group used a manual puzzle for the traditional learning method and the experimentation group used the augmented reality application. The results displayed that the experimentation group carried out the activities faster and performed better than the control group.

In \cite{lozano2018tangible} the authors present a novel interactive system whose aim is for students with visual impairment to be able to learn basic concepts with the help of this interface. This system does not have any visual interface since the users mainly have to interact with touch and speech recognition. The students use a tangible user interface where they have to grasp physical objects to complete the goals in this serious game. These objects have NFC tags which are identified by the mobile phone with an installed NFC reader which makes it possible to recognize the different objects. Another important factor for interaction is audio, this being the way to communicate with users who receive instructions and then provide feedback when being guided through the learning process. According to the authors positive results were achieved during the assessment of the usability and convenience of the system.

\begin{table}
	 		\centering 
	 		\caption{Comparing different studies and this work. (NI: Natural Interaction / UM: User Model / IA: Interaction Adaptation / GR: Gesture Recognition / VI: Visual Impairment / HI: Hearing Impairment / PI: Physical Impairment / A: Autism / AM: Adaptation Mechanism).}
\label{table_Comparison}
\begin{turn}{90}
{\fontsize{6.30}{6.5}\selectfont

\begin{tabular}{p{2,5cm}p{1,7cm}p{1,7cm}p{1,7cm}p{1,7cm}p{1,7cm}p{1,7cm}p{1,7cm}p{1,7cm}p{2.75cm}}
\hline\noalign{\smallskip}
Projects & NI & UM & IA & GR & VI & HI & PI & A & AM \\
\noalign{\smallskip}\hline\noalign{\smallskip}
 \raggedright Kinems \cite{altanis2013children,cai2018case}& 
 Kinect & 
 None & 
 None & 
 \raggedright Hand and body gestures & 
 None & 
 None & 
 \raggedright Activities designed for physical impairment & 
 \raggedright Activities designed for autism & 
 None
 \\
 \noalign{\smallskip}\hline\noalign{\smallskip}
 Cai et al \cite{puspitasari2013kidea} & 
 Leap Motion & 
 None & 
 None & 
 \raggedright Hand gestures & 
 None & 
 None & 
 None & 
 \raggedright Activities designed for autism & 
 None
 \\
 \noalign{\smallskip}\hline\noalign{\smallskip}
 \raggedright Hsiao et al \cite{hsiao2016using} & 
 \raggedright ASUS Xtion PRO & 
 None & 
 None & 
 \raggedright Body gestures & 
 None & 
 None & 
 None & 
 None & 
 None
\\
\noalign{\smallskip}\hline\noalign{\smallskip}
Topolor \cite{shi2013topolor,shi2013social,shi2013evaluation} & 
None & 
\raggedright Overlay user model & 
None & 
None & 
None & 
None & 
None & 
None & 
Learning Concept \par Adaptation Strategy + \par Learning Path \par  Adaptation Strategy + \par Learning Peer \par Adaptation Strategy
\\
\noalign{\smallskip}\hline\noalign{\smallskip}
Ghiani et al \cite{ghiani2015dynamic} & 
\raggedright Neurosky headset + Tobii Eye Tracker & 
\raggedright User Model Manager & 
None & 
None & 
None & 
None & 
None & 
None & 
Adaptation engine + \par Adaptation rules
\\
\noalign{\smallskip}\hline\noalign{\smallskip}
TECH8 \cite{dolenc2015tech8} & 
None & 
None & 
None & 
None & 
None & 
None & 
None & 
None & 
Intelligent and \par adaptive module + \par metadata
\\
\noalign{\smallskip}\hline\noalign{\smallskip}
UZWEBMAT \par 
\cite{ozyurt2012integrating,ozyurt2013design,ozyurt2013integration,ozyurt2014effects} & 
None & 
None & 
None & 
None & 
None & 
None & 
None & 
None & 
Expert system
\\
\noalign{\smallskip}\hline\noalign{\smallskip}
Cinquin et al \cite{cinquin2020designing} & 
None & 
None & 
None & 
None & 
None & 
None & 
None & 
None & 
Framework for accessibility
\\
\noalign{\smallskip}\hline\noalign{\smallskip}
\raggedright ATHYNOS \cite{avila2018athynos} & 
\raggedright Kinect & 
None & 
None & 
Body gestures & 
None & 
None & 
\raggedright Activities designed for physical impairment & 
None & 
None
\\
\noalign{\smallskip}\hline\noalign{\smallskip}
Lozano et al \cite{lozano2018tangible} & 
\raggedright Tangible user interface + \par speech recognition & 
None & 
None & 
None & 
\raggedright Activities designed for visual impairment & 
None & 
None & 
None & 
None
\\
\noalign{\smallskip}\hline\noalign{\smallskip}
\raggedright Faisal et al \cite{faisal2016innovating} & 
\raggedright Kinect & 
None & 
None & 
Body gestures & 
None & 
None & 
\raggedright Activities designed for physical impairment & 
None & 
None
\\
\noalign{\smallskip}\hline\noalign{\smallskip}
{\bf Our proposal} & 
Kinect & 
\raggedright Feature-based user model & 
\raggedright Device-interaction model & 
\raggedright Body gestures & 
\raggedright Activities designed for visual impairment & 
\raggedright Activities designed for hearing impairment & 
\raggedright Activities designed for physical impairment & 
\raggedright Activities designed for autism & 
Rule-based system + \par device-interaction model + user model
\\
\noalign{\smallskip}
\hline
\end{tabular}
}
\end{turn}
\end{table}

Table \ref{table_Comparison} shows the features of each of the aforementioned studies as well as the work developed in this paper.

\section{Device-Interaction for Special Needs} \label{proposal}

The main feature of the proposed system is its adaptability with regard to interaction, this being a fundamental aspect in any computer application. This work was designed for users with physical and sensory disability because the aim is to improve interaction according to the user characteristics and ensure an optimal experience. Two experiments were carried out to evaluate the validity of the hypothetical idea. The first experiment was an assessment by experts in which the Cognitive Walkthrough with Users technique was applied. The experts included an interactive expert, an educator who specializes in special education and two teachers from the special education center.

Afterwards, an experiment with twelve students from the special education center was organized. There were students with autism, physical disability, visual impairment and severe hearing loss. They had to complete the activities developed in this work and the time and the errors that they made during the performance were recorded. There are two types of activities; an activity to associate concepts about a topic and another activity to work on the learners laterality. The concept of association laterality has two versions: the animal theme and the vehicle theme. In this proposal, natural interaction was integrated because it was more convenient for special needs students since the only requirement for interaction with the device is that the user had to be within a certain distance from this device. The Microsoft Kinect v2 sensor was included due to the interaction requirements since this device can recognize body motion. The necessary elements for carrying out an adaptable interaction are: user model, device-interaction model and the adaptation rules.

\subsection{User Model} \label{sub_UserModel}

The user model designed in this work is a features-based model \cite{brusilovsky2007user}. The tutors register their students in the system, filling in the necessary data for the student user model. This model stores the following user characteristics: full name, age, sex, laterality problems and disability.

In this work the most relevant features are: laterality problems and disability. When we talk about disability, it is necessary to mention the International Classification of Functioning, Disability and Health by World Health Organization (WHO)\footnote{Int. Classification of Functioning, Disability and Health - \url{https://bit.ly/3h7AOUO}},
whose main objective is to establish a unified language for the description of health and its states where disability is dealt with, among other items \cite{cieza2008international}. The disabilities reflected by the user model are included in this classification and they are: people with severe hearing loss, visual impairment, physical disability or autism. These features have been included because of the type of users who participated in this research. The laterality problems feature describes when the users cannot distinguish between their left and right side of their bodies perfectly.
From this model it should be pointed out that the disability features and laterality problems have an important weight in the adaptation process as they will affect the device-interaction model.

\subsection{Device-interaction Model} \label{sub_Dev-Int_Model}

The device-interaction model takes into account the user characteristics, but this model focuses especially on the device. This is because the device is the means by which the students are able to use the system and this is why it has an important role in this model. The aim of this model is to optimize user interaction with the system by virtue of the inherent features of the device. The device-interaction model characteristics are:

\begin{itemize}
\item \textbf{Detection of the biped position}:  This property determines if the user is  standing or sitting. This system will be used by students confined to a wheelchair and detecting this is fundamental for the activity adaptation. When it is detected that the user is sitting, a more meticulous tracking of the upper-limbs is carried out while the lower-limbs are ignored. Nevertheless, if it is detected that the user is standing, the 26 joints recognized by Microsoft Kinect v2 are taken into account.
\item \textbf{RGB camera activation}: Some of the students have a very low cognitive level and they are not able to associate their movements with the screen elements unless they can see themselves on the screen. This feature makes it possible to activate the RGB camera from the device so that the students can identify themselves on the screen and interact with the environment.
\item \textbf{The depth distance}: This attribute saves the distance the user has to be from the Kinect device in order to interact properly. However, there is a recommended user distance for this sensor (1.2 - 3.5m) depending on the user’s height.
\item \textbf{Arm motion}: This feature identifies whether the user can move both arms or only one of them. It is necessary to move at least one in  order to interact with the system. This attribute is for students who have some physical disability and can only move one of their arms. In the case that the user can use both arms, it will determine the users dominant arm to make the interaction easier regardless of whether the user is left-handed or right-handed.
\end{itemize}

\subsection{Adaptation rules} \label{sub_Rules}

The adaptation rules are essential for adapting the activities to the user characteristics. The rules in the project were initially based on Accessible Games Standard v1.0\footnote{Accessible Games Standard v1.0 - \url{https://bbc.in/31Wx2Hi}}
and Game Accessibility Guideliness\footnote{Game accessibility guideliness - \url{http://gameaccessibilityguidelines.com/}} \cite{garber2013game}. This is due to the fact that the activities in the project included game-based mechanics and Kinect is a tool that was designed to interact with games. However, we made some improvements in the adaptation rules after receiving some feedback from experts in special needs education who participated in the study. These rules are defined {\em a priori} and are triggered depending on the user model and device-interaction model features. The most important rules are explained below. Table \ref{table_Rules} summarizes the rules created for this study. It shows the actions or changes that the system makes when it detects whether the user has a visual impairment, severe hearing loss or physical disability, autism, motion limitation in his/her upper-limb or if he/she is wheelchair user.

\begin{table} [!ht]
\renewcommand{\arraystretch}{1.3}
\caption{Adaptation Rules. (I: Instructions / BC: Background color / 3DC: 3D objects color / IM: Interaction Mode / Fed: Feedback / G: Gestures / SVI: Show visual icons / D: Distance between elements / MD: Motion detection).}
\label{table_Rules}
\centering
\begin{turn}{90}
{\fontsize{8}{10}\selectfont
\begin{tabular}{p{0.15cm}p{3.75cm}p{0.5cm}p{0.5cm}p{0.75cm}p{1.25cm}p{1.5cm}p{0.25cm}p{0.25cm}p{1cm}p{2cm}}
\hline\noalign{\smallskip}
\# & Disability & I & BC & 3DC & IM & Fed & G & SVI & D & MD  \\
\noalign{\smallskip}\hline\noalign{\smallskip}
1 &  Visual & Audio & Black & Yellow & Collision &  Audio & No & No & Standard & Dominant arm \\
2 &  Hearing & Visual & Image & Normal  & Gestures & Visual & Yes & No & Standard & Dominant arm \\
3 &  Physical & Audio & Image & Normal & Collision & Visual\&Audio & No & No & Standard & -  \\
4 &  Autism & Audio & Image & Normal & Drag\&Drop & Visual\&Audio & No & Yes & Standard & Dominant arm  \\
5 &  Physical (wheelchair) & Audio & Image & Normal & Collision & Visual\&Audio & No & No & Reduced & Dominant arm\\
6 &  Physical (mov. right arm) & Audio & Image & Normal & Collision & Visual\&Audio & No & No & Standard & Right arm\\
7 &  Physical (mov. left arm) & Audio & Image & Normal & Collision & Visual\&Audio & No & No & Standard & Left arm\\
8 &  Physical (mov. both arms) & Audio & Image & Normal & Collision & Visual\&Audio & No & No & Standard & Dominant arm  \\
\noalign{\smallskip}\hline
\end{tabular}
}
\end{turn}
\end{table}

Rule \#1 applies to the action carried out by users with visual impairment. In this case, all the activity instructions are audio, the background is colored black and the 3D objects that the user interacts with are yellow to contrast with the background and make them easier to identify. The interactive mode is based on tracking either the arm that the user can move or the one previously chosen. Using collision means that when the user has to interact with the elements in the scene, the collision between the cursor controlled by the user and the 3D objects will be detected. The activity feedback is audio so that the user knows when he or she has made a mistake.


Rule \#2 is for users with severe hearing loss. In this situation, all the activity instructions are visual and the interaction mode is based on gesture recognition. The without delay mode in the gesture recognition means that there is a time variable which measures the time from the starting position to the final one. In this case, this variable is assigned to the minimum value in which the gesture is identified immediately so that the user does not feel frustration by having to wait for the answer from the system to his or her action. Finally, the feedback is in visual mode.

Rule \#3 refers to the actions carried out by users with physical disability. In this case, all the activity instructions are audio. The instructions could have been in visual mode because this user profile does not have any sensory disability, but this mode was chosen instead because it is faster for the students to listen to the instructions than to read them. The interactive mode is based on collision motion detection and the feedback related to the activity is both audio and visual.

Rule \#4 refers to users with autism. In this situation, all the activity instructions are in audio for similar reasons to the previous rule. The interaction mode is based on drag and drop motion detection. In this type of interaction, when the collision between the cursor and one element is detected, this element is dragged until it reaches an area where the user can drop it. Some visual icons are shown next to the 3D objects to assist with students’ comprehension since the students with autism are used to working with pictograms. The feedback is audio and visual.

Rule \#5 applies to users who are wheelchair-bound. In this case, the action is to reduce the distance between the elements because in this position, the movement is more limited than when the user is standing. In this way, the user will be able to interact with the different interface elements more easily.

Rule \#6 is for users that have a physical disability that affects their left arm motion while Rule \#7 is for those with physical disability that affect their right arm motion. In this situation, the joints detected for the tracking and detect the motion are related to the left arm for the former and the right arm for the latter.

Rule \#8 is created specifically for users with physical disability who can move both arms. This rule provides them with the option to use either one (usually the dominant arm) for greater comfort during the interaction.

\subsection{Adapted System with a Device-Interaction Model} \label{sub_Architecture}

\begin{figure}
\centering
\includegraphics[width=4.15in]{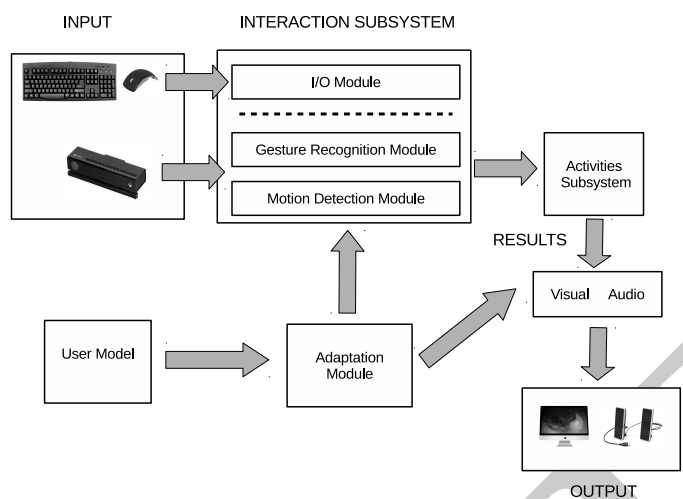}
\caption{System Architecture.}
\label{figArchitecture}       
\end{figure}

The design supports different types of input devices which interact in the system (see Figure \ref{figArchitecture}). These devices are mouse, keyboard and Microsoft Kinect v2. This architecture is composed of distinct subsystems that make this system adaptable according to the user features, especially with regard to interaction. Here, the functionality of each system component is described.

The interaction system has three components: I/O Recognition, Gesture Recognition and Motion Detection. The usage of each component depends on the device which the user is using for interaction. For example, when the tutor is organizing profiles or signing in a new user, the input devices are the mouse and the keyboard since this task is more difficult to complete with the Kinect v2 device. However, when a student is doing an activity, he or she uses the Kinect v2 sensor as it is more suitable for this task. The I/O Recognition component is activated when the mouse and keyboard are used and the Kinect v2 device is excluded in this process in order for the interaction to work smoothly. The components called Gesture Recognition and Motion Detection can only be activated when the Microsoft Kinect v2 is used for the interaction. However, the decision to activate each component is made by the Adaptation Module. The Adaptation Module comprises a set of rules based on the user model in order to identify the user characteristics and offer a suitable interaction mode for the user.

The Gesture Recognition module identifies two gestures; raising the left arm and raising the right arm. In the initial phase it was decided to integrate these basic gestures to make the execution easier for the students (especially for those ones with physical disability). The gesture recognition process is carried out by a state finite machine where each state checks if the current pose from the user is identical to the one stored in that state. The states will have an initial and ending state, where it will check the initial and final pose of the gesture for recognition and a set of intermediate states that will evaluate the gesture to verify that this gesture coincides with the one the user wants to be recognized. Evaluating only the initial and final position is insufficient because the user is able to perform the initial gesture (a circle movement) and continue to the final position when what the system wants to recognize is a bottom-up movement.

The aim of these intermediate states is to check that the movement trajectory is correct throughout the gesture validation process. In these gestures the right hand, left hand and left shoulder are considered. To validate the initial position, the hand position made by the left shoulder has to be checked. The final position validation takes into consideration two aspects: the hand position with respect to the body is verified and second, that the gesture is made within a fixed time, for example, if it takes a long time for the user to get from the initial position to the final position, the gesture would not be valid.

As soon as the gesture recognition module is activated, the gesture listener will begin as well. Its function constantly checks if the user meets the criteria of the initial state of some of the gestures to complete the gesture recognition process and inform the system if the gesture has been made correctly.

The Detection Motion Module uses the Microsoft Kinect v2 Software Development Kit (SDK)\footnote{Kinect Windows SDK: \url{https://msdn.microsoft.com/en-us/library/dn799271.aspx}}: a development kit for creating applications that support the Kinect technology to assign one of the 25 joints or limbs which it recognizes and so it  will track only those specific joints. The joint or limb will be selected depending on the Adaptation Module, which is responsible for informing the Detection Motion Module about which joint or limb to track. This module is working with the left and right hand because its goal is to move a hand-shaped 3D object as a cursor and in this way interact with other 3D objects that will be shown on the screen. Therefore, in order for the hand to behave as a cursor and to make the 3D object moves in unison with the user hand movement, the X, Y and Z coordinates from both of them are matched.

Besides the Interaction Subsystem, the proposed environment has an Activities Subsystem, to organize the activities that the user does according to his or her profile and these are adapted to his or her characteristics. This system takes into account two user model features: the disability and laterality problem. If the user does not have any laterality problem, the system does not provide this option for the user to select it in the interface. The Adaptation Module gets the information from the user models, device-interaction model and the associated rules that are sent to the Activities Definition module, which is responsible for creating activities that enable the instructions, feedback, components, logic, and interaction to have specific characteristics depending on the user disability. Furthermore, this module also enables the Interaction Model to make use of the Kinect v2 device characteristics according to the preferences that have been saved from the automatic verification process. In this way, besides organizing the activity according to the type of user disability, these three features are also considered: the common user posture (wheelchair or standing), the RGB camera activation or the arm used for the interaction.

The Environment component results show a feedback to the users depending on their actions in the system. The feedback might be visual or audio. The visual feedback is represented with a smiley face or a sad face and the audio with a particular sound from the object which has been selected or a sound associated with an error. In this process, it is also checked whether the user selection corresponds to the activity request. A different type of feedback would be given to the user depending on whether the action is correct or wrong. The Adaptation Module is also a part of this component as, depending on the type of disability, the feedback is shown in a different way. For example, if the student has severe hearing loss, the feedback will be visual, whereas if the learner has visual impairment, the feedback will be audio.

\subsection{Proposed Interactive Activities} \label{sub_Activities}

In this section, the implemented activities (see Table \ref{table_ActivityFeatures}) in the prototype are described and each specification relating to the user characteristics is explained in detail. Two types of activities were developed. The teachers proposed the activities that were useful for their students at that time and they guided us during the development phase. Their students found it hard to differentiate between some concepts and their hand, eye, foot, or ear dominance were not consistently right- or left-sided (crossed laterality). This crossed laterality was discovered around forty years ago and affects the organization of the upper functions in our system, this disorder affects language and mathematics learning, analytic, logical, understanding and concentration skills, time-space perception and balance, among others \cite{ferrero2017crossed}. Thus, the goal of one activity is for the student to associate concepts and the other one is to improve the laterality issue. The user can choose between two topics in the concept association activity: animals or vehicles. From there, the tutor will select one of them depending on what he or she wants to teach on that occasion.

\begin{table*}[!t]
\renewcommand{\arraystretch}{1.3}
\caption{Summary about the Main Features of the Activities.}
\label{table_ActivityFeatures}
\centering
\begin{turn}{90}
{\fontsize{7.25}{9}\selectfont
\begin{tabular}{lll}
\hline\noalign{\smallskip}
{\bf Mode} & {\bf Concepts Association} & {\bf Laterality Activity}  \\
\noalign{\smallskip}\hline\noalign{\smallskip}
\multirow{5}{*}{Visual} &  \tabitem Instructions: Audio & \tabitem Instructions: Audio \\
 & \tabitem Background color: Black & \tabitem Background color: Black \\
 & \tabitem 3D objects color: Yellow & \tabitem 3D objects color: Yellow \\
 & \tabitem Interaction: Motion detection & \tabitem Interaction: Motion detection \\
 & \tabitem Feedback: Audio & \tabitem Object translation \\
 \hline
\multirow{4}{*}{Hearing} &  \tabitem Instructions: Visual & \tabitem Instructions: Visual \\
 & \tabitem Interaction: Gesture recognition & \tabitem Interaction: Gesture recognition \\
 & \tabitem Feedback: Visual & \tabitem Feedback: Visual \\
  & & \tabitem Object shift (left / right) \\
 \hline
\multirow{5}{*}{Physical} &  \tabitem Interaction: Adaptation according to the user mobility & \tabitem Instructions: Audio \\
 & \tabitem Elements position according to the movement & \tabitem Interaction: Movement detection \\
 & \tabitem Feedback: Visual and audio & \tabitem Shift interval according to movement \\
 &  & \tabitem Location borders on the screen \\
 &  & \tabitem Feedback: Visual and audio \\
 \hline
\multirow{6}{*}{Autism} &  \tabitem Instructions: Visual and audio & \tabitem Instructions: Audio \\
 & \tabitem Interaction: Drag \& drop & \tabitem More number of interaction elements \\
  & \tabitem Elements identification through pictograms & \tabitem Objects shift \\
 & \tabitem Feedback: Visual and audio & \tabitem Identification through pictograms \\
 &  & \tabitem Interaction: Drag \& drop \\
 & & \tabitem Feedback: Visual and audio \\
\noalign{\smallskip}\hline
\end{tabular}
}
\end{turn}
\end{table*}

\subsubsection{Activity about Concept Association}

\begin{figure*}[!h]
\centering
\subfloat[]{\includegraphics[width=2.1in]{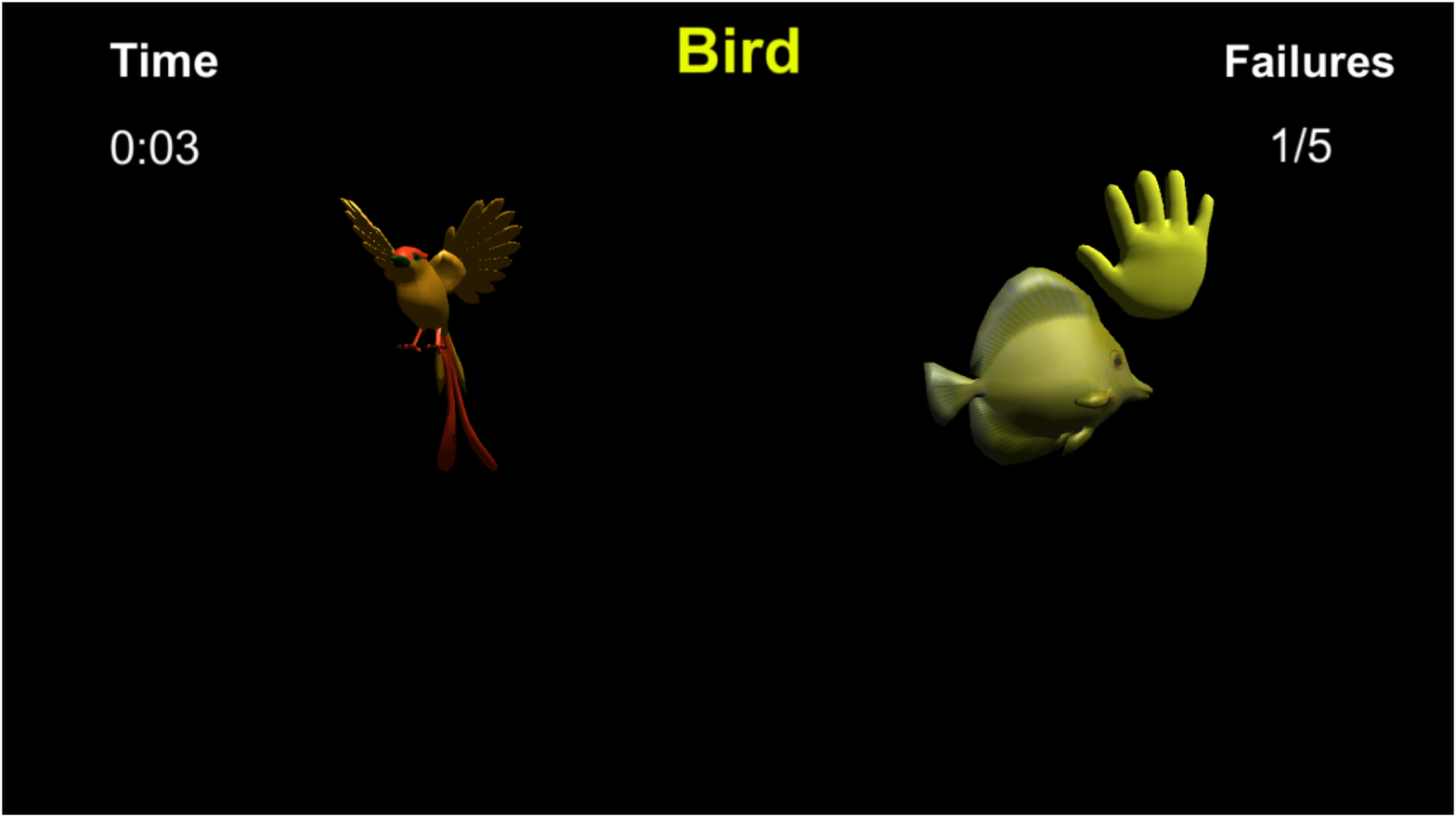}%
\label{f:visualConceptAct}}
\hfill
\subfloat[]{\includegraphics[width=2.1in]{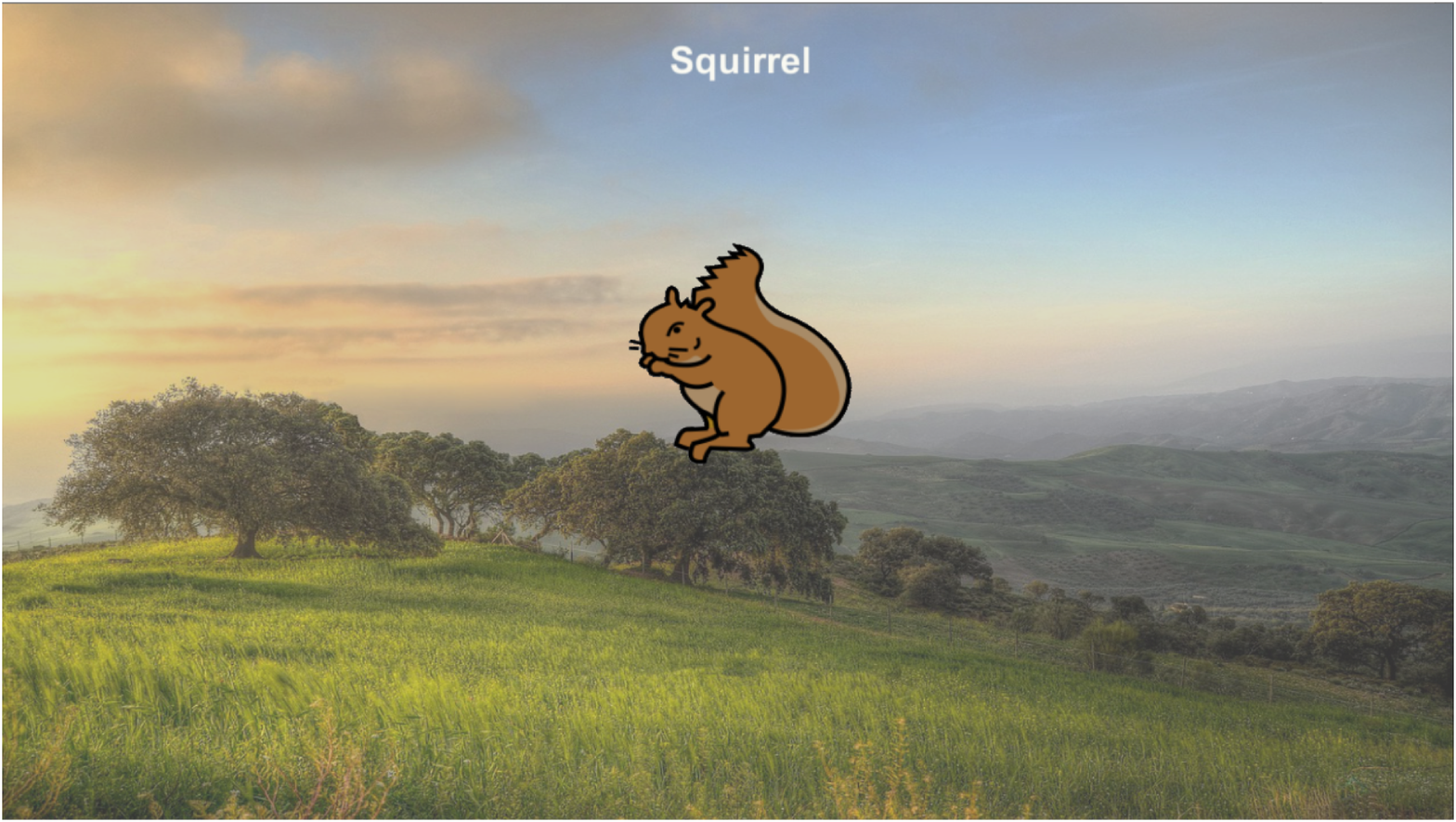}%
\label{f:insAudConceptAct}}
\hfill
\subfloat[]{\includegraphics[width=2.1in]{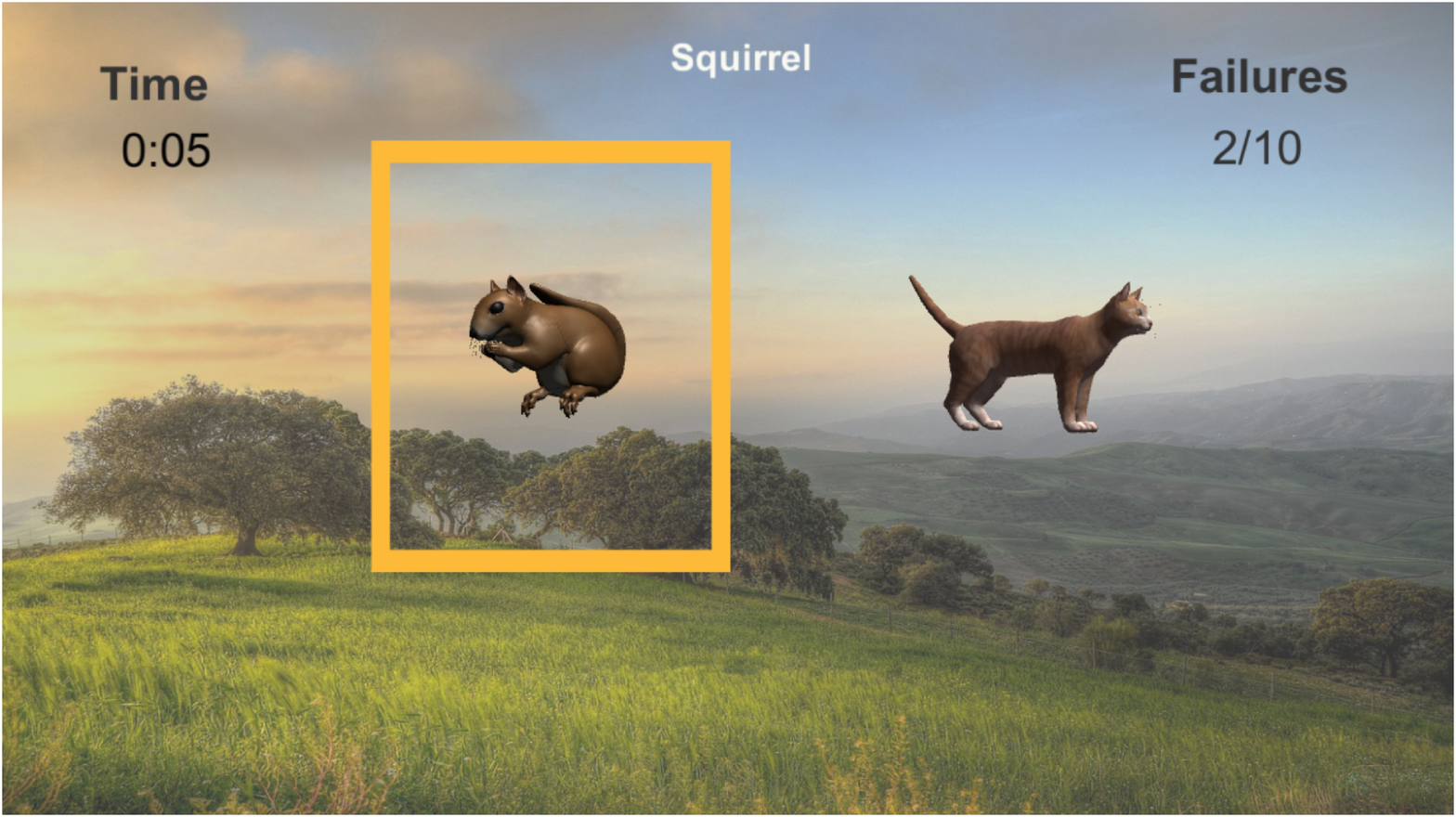}%
\label{f:audConceptAct}}
\hfill
\subfloat[]{\includegraphics[width=2.1in]{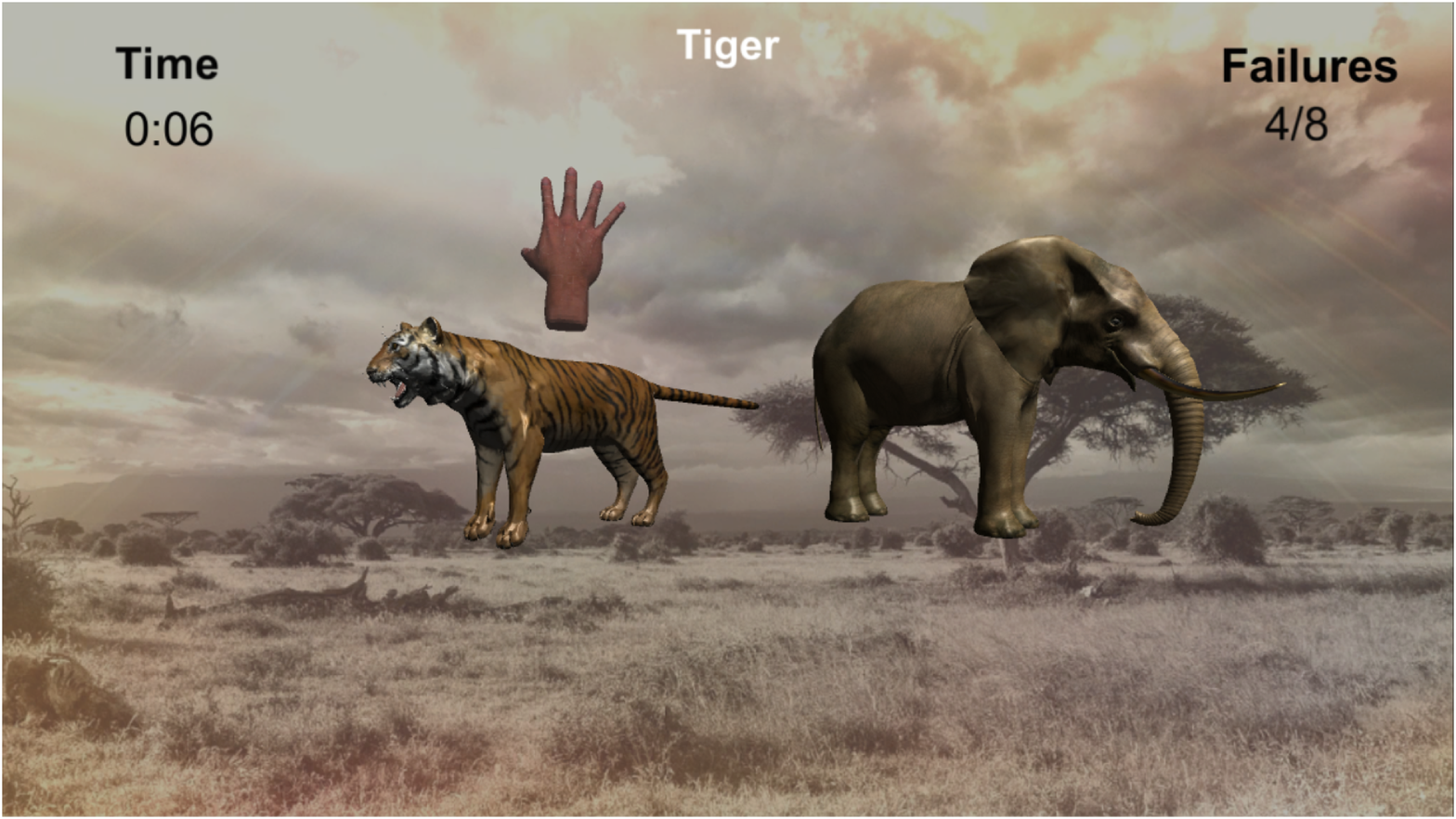}%
\label{f:physConcepActiviy}}
\hfill
\subfloat[]{\includegraphics[width=2.1in]{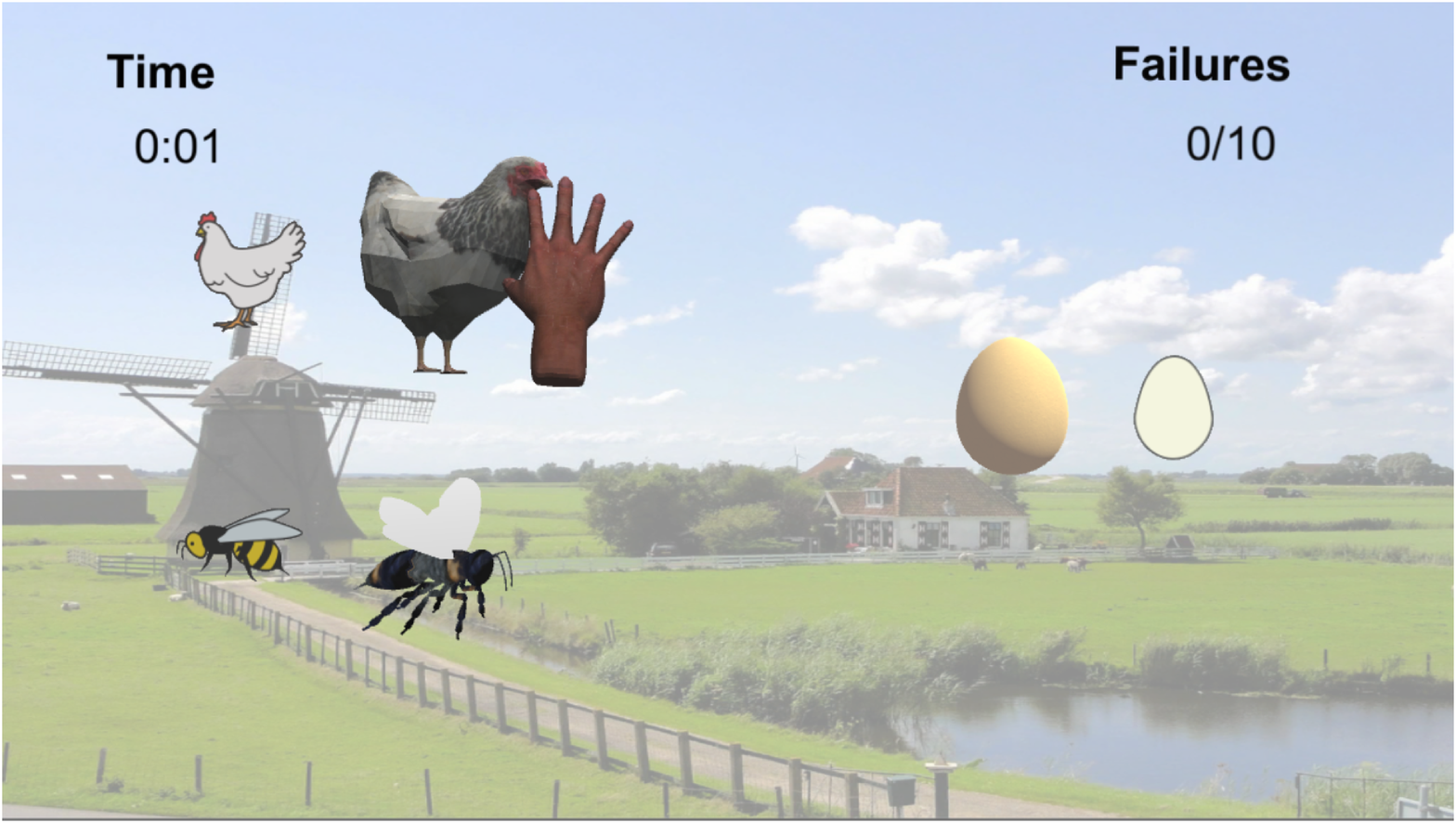}%
\label{f:autismConceptActEgg}}
\hfill
\subfloat[]{\includegraphics[width=2.1in]{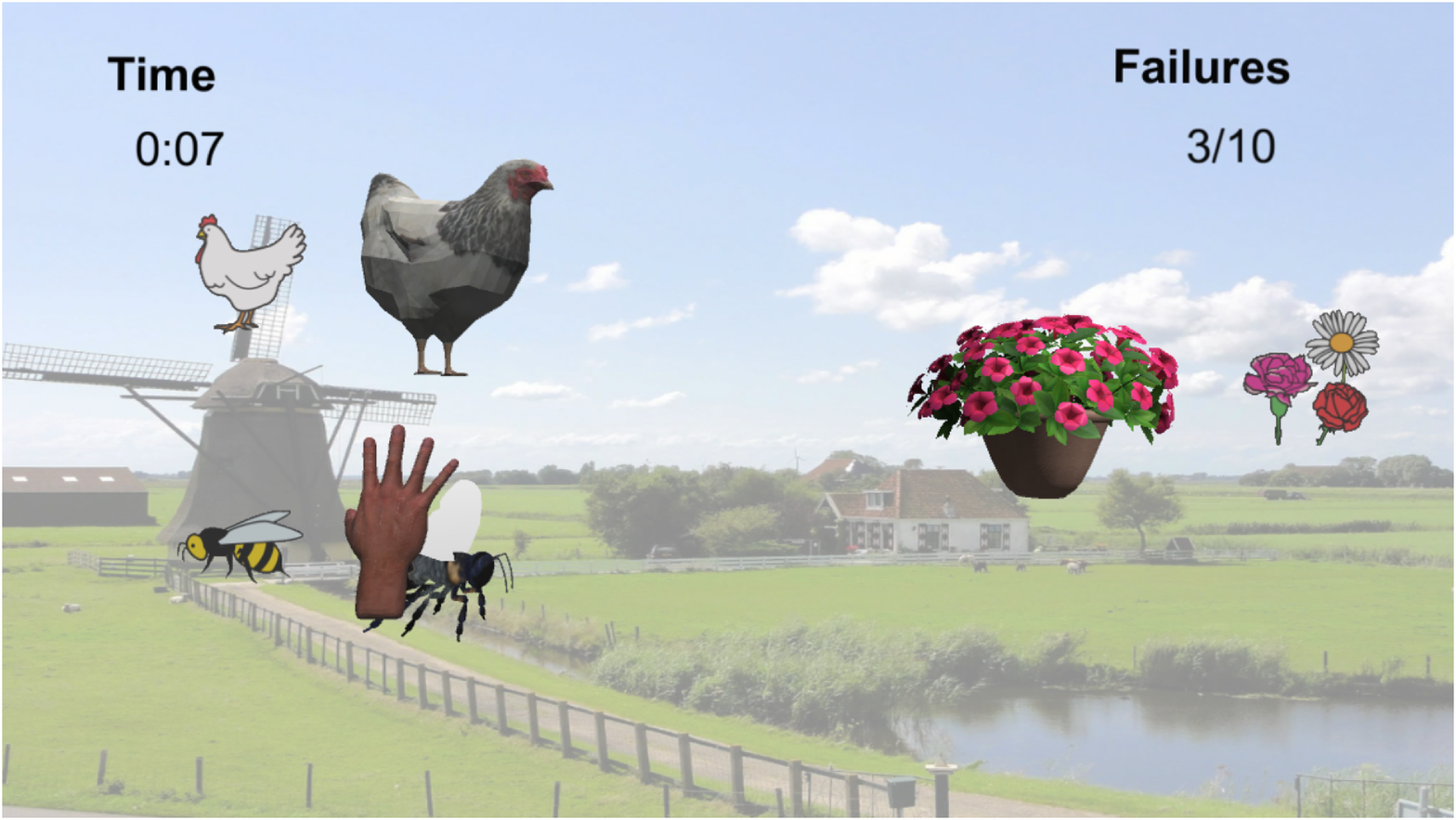}%
\label{f:autismConceptActFlowers}}
\caption{Screens Related with the Concept Association Activity. a) Contrast between the elements and the background for visual impairment students; b) The previous screen before starting the activity for hearing impairment students; c) The option chosen is selected; d) The objects are closer for physical impairment cases; e) and f) Activities for users with autism with different goal.}
\label{screensConceptAct}
\end{figure*}

The aim of this activity is for the students to learn concepts within a theme (see Figure \ref{screensConceptAct}) and the system provides the essential information for the student to identify the model which he or she has to select. This information is shown in a visual way, or relayed in audio according to the user impairment. When the student has selected one of the options, it will allow the user to modify them within a certain time period. Then, the system will give positive or negative feedback to the user depending on the user action. This sequence can be repeated a number of times according to what the tutor deems appropriate. The means of interaction is different depending on the user model and the device-interaction model. The different versions according to the user disability are described next.

\begin{itemize}
\item \textbf{Visual impairment}: In this case the GUI colors are changed: the back- ground color is black and the 3D object color is yellow to create contrast (see Figure \ref{f:visualConceptAct}). The instructions shown at the beginning of the activity are audible. The interaction is through the motion detection, where the users can control a hand-shaped cursor according to their dominant arm. When one of the elements is selected, a yellow frame will appear around the element selected in order to know which element has been selected by the user. The feedback is exclusively audio.

\item \textbf{Hearing impairment}: In this version, the elements and the interface colors are shown without any modification (see Figure \ref{f:audConceptAct}). The instructions are shown in a visual way (see Figure \ref{f:insAudConceptAct}) and the interaction is through gesture recognition. The gestures are raising the right arm and the left arm. If the left arm is raised, the element located at the left side of the screen is selected, but if the right arm is raised, the element placed on the right side will be selected. The feedback on this version is totally visual. If the answer is correct, a smiley face will appear on the screen, otherwise a sad face is shown.
\item \textbf{Physical Impairment}: When the user has a physical disability, the changes relating to the interaction will depend on the values in the device-interaction model: whether the person is using their left or right arm. For example, if a person can only move their right arm, the system will select that arm to interact with the different elements in the interface. Another relevant aspect is that if the user is confined to a wheelchair, the elements will be closer and centered on the screen (see Figure \ref{f:physConcepActiviy}) so that the movement range is reduced, making it more accessible for the user. The feedback is visual and audio, unlike the previous situation.
\item \textbf{Autism}: The instructions are shown in a visual and audio form. The mode of interaction with the elements is distinct because it follows a methodology that  is more complicated such as drag \& drop. The aim is to associate the elements on the left side of the screen with the ones which are placed on the right side correctly (see Figures \ref{f:autismConceptActEgg} and \ref{f:autismConceptActFlowers}). For this reason, when the cursor collides with an element located on the left side, this element is dragged by the cursor through tracking and when it makes contact with the right side element, the tracking process is stopped. There are some pictograms next to the different objects in order to make it easier for them to identify each element. The feedback is visual and audio.
\end{itemize}

\subsubsection{Laterality Activity}

The goal of this activity is to work the laterality from left to right. At the beginning of the activity, a ball appears in the middle of the screen and the user has to touch it with the cursor. When the cursor and the 3D object collide with each other, the ball will be moved on the screen. This translation depends on the laterality problem of each student: If the user does not recognize the right side then the object will be moved a few centimeters to the right but if the user is not able to recognize the left side, then the object will be moved a few centimeters to the left. This sequence will be repeated a number of times until the ball reaches a default position on the screen. The way of interaction with the system depends on the user characteristics:

\begin{itemize}
\item \textbf{Visual impairment}: The instructions are in audio. This activity provides color contrast like the previous activity in which the background is painted with black while the ball is yellow. The interaction of the activity is motion detection, where the user has to make contact between the cursor which is controlled by the dominant user arm and the ball. Once the collision is made, this ball is moved to the corresponding direction depending on whether the user has problems with right or left laterality.
\item \textbf{Hearing impairment}: The instructions are visual. The user interacts with the system through gesture recognition. When the raising the right arm gesture is identified, the ball is moved to the right if the option right laterality was selected. Otherwise, if raising the left arm is recognized, the ball is moved to the left if the option left laterality was selected.
\item \textbf{Physical impairment}: The interaction is the same as in the visual impairment version, with the exception that the background is not black and the ball is colored yellow to provide contrast between the elements. When the user is in a wheelchair, the 3D model does not move to the other end of the screen and the shift is lower than in other versions so that the user does not have to make any awkward movements to complete the activity.
\item \textbf{Autism}: At the beginning, the instructions for the activity are communicated through audio. In this version, the ball always starts at the center of  the screen and a 3D basketball basket is moved instead of the ball. When the user causes a collision between the cursor and the ball, the user can drag the ball until it reaches the basket. The basket is moved to the corresponding direction depending on the version that has been selected.
\end{itemize}

\section{Experiments and Results} \label{results}

This section describes the procedure to assess the system. There are two types of evaluation involved: an evaluation by experts and an evaluation with end users. In the evaluation by experts, a technique called Cognitive Walkthrough with Users is used in a combination with the Thinking Aloud method, in which two experts and two teachers participated. In the evaluation with the end users, students with different types of disability from the Special Educational Needs Center Princesa Sofia tried the prototype. Certain factors that were considered relevant for this study were measured to obtain a conclusion about this experiment.

\subsection{Expert Evaluation}

The evaluation techniques are classified in different categories depending on the method’s features. The general classification is: indagation, inspection and test \cite{granollers2006incorporation}. The inspection method is the most interesting one for this part of the experiment since evaluators (experts)  give  their  judgment  regarding  the system’s usability and accessibility  \cite{kushniruk2015integrating}. The  Cognitive  Walkthrough is the inspection method that it is included in this assessment because it is suitable for interactive adaptive systems \cite{dhouib2016classification} and its several advantages: it is inexpensive, it can be carried out at an early stage of development and it requires little effort, one expert is usually enough. On the other hand, very detailed analysis of the tasks is required and there is no rating by ”real” users \cite{dhouib2016classification,lewis1997cognitive}.

In this evaluation user collaboration is fundamental to this system and thus, two expert evaluators in interactive system usability and special needs participated in this experiment together with two teachers from the center of special education. In this phase of the evaluation, Cognitive Walkthrough was used together with a method called Thinking Aloud. The inclusion of this technique has rendered this evaluation process more thorough since the users have to express their thoughts while simultaneously interacting with the system. In addition, the participants can be asked about the reasons they are carrying out a certain action in situ or if they consider some aspects of the prototype confusing. From this evaluation, some important information was extracted with which to improve the graphical interface and the functionality of the prototype with the purpose of making its implementation easier and more intuitive for the final user.

In order to conduct the experiments using Cognitive Walkthrough with Users, you must first do the Cognitive Walkthrough in the traditional way and after this, the users are incorporated into the study \cite{granollers2006incorporation}. With this premise, this method was applied first by experts in interactive system usability and special needs. Afterwards, the two professors carried out the Cognitive Walkthrough in combination with Thinking aloud since this technique is usually applied to end users of the system. At this point it is necessary clarifying that the session with every participant was individual. It is common that a low-fidelity prototype is used in the Cognitive Walkthrough and for Thinking aloud a high-fidelity prototype \cite{zaini2019evaluation}. However, we made the decision of using a high-fidelity prototype in this evaluation because both techniques were combined and the goal was that all the users had the same prototype to carry out this evaluation. During the Cognitive Walkthrough, the evaluators took notes while they performed the tasks on the negative aspects or those that needed to be improved. The teachers were recorded in audio since the Thinking aloud consists of saying what they think about the system at that time, but they were also allowed to take notes in case they considered it appropriate to write aspects that we should take into account when improving the final prototype.

The experts completed a set of tasks pertaining to the system. This system is a way for each tutor to manage the student information that is stored in a database. They were asked to complete several tasks that could be divided by two categories: Management and Activities. The Management category  would enclose the following tasks:
\begin{itemize}
    \item Register a teacher
    \item Login in to the system
    \item Add a student
    \item Access to the menu called My profile
    \item Edit the profile data
    \item Select one student from the list
    \item Edit the student data
    \item Delete a student profile
    \item Go back to the students' list
    \item Go to the activities selection screen
    \item Select an activity
    \item Log out
\end{itemize}

And in the Activities category the participants had to complete the activities successfully and fail as well to check the distinct feedback. These tasks would be:
\begin{itemize}
    \item Complete the activity for people with autism
    \item Complete the activity for people with severe hearing loss
    \item Complete the activity for people with physical disability
    \item Complete the activity for people with visual impairment
\end{itemize}

From this evaluation, some results were analysed to improve the experiment:
\begin{enumerate}
\item[(a)] Modifications in the GUI, for instance, adding icons in some buttons instead of text or changing the location of some elements in the graphical interface. 
\item[(b)] Implanting the gesture recognition system for the users with severe hearing loss because they are used to communicating with this type of interaction, instead of using the motion detection.
\item[(c)] Suggestion about the 3D elements size for the students with visual impairment.
\item[(d)] The use of pictograms for students with autism so that they could comprehend the task better.
\item[(e)] The creation of a laterality activity.
\item[(f)] In the association concept activity it is necessary for the option selected to be framed.
\item[(g)] The system should not play music during the task in order to avoid distracting the student.
\item[(h)] The aim of the activity has to be highlighted in such a way so that the students attention will not be distracted by the background.
\end{enumerate}

\subsection{User Evaluation}

\begin{figure*}[!h]
\centering
\subfloat[]{\includegraphics[width=2.2in]{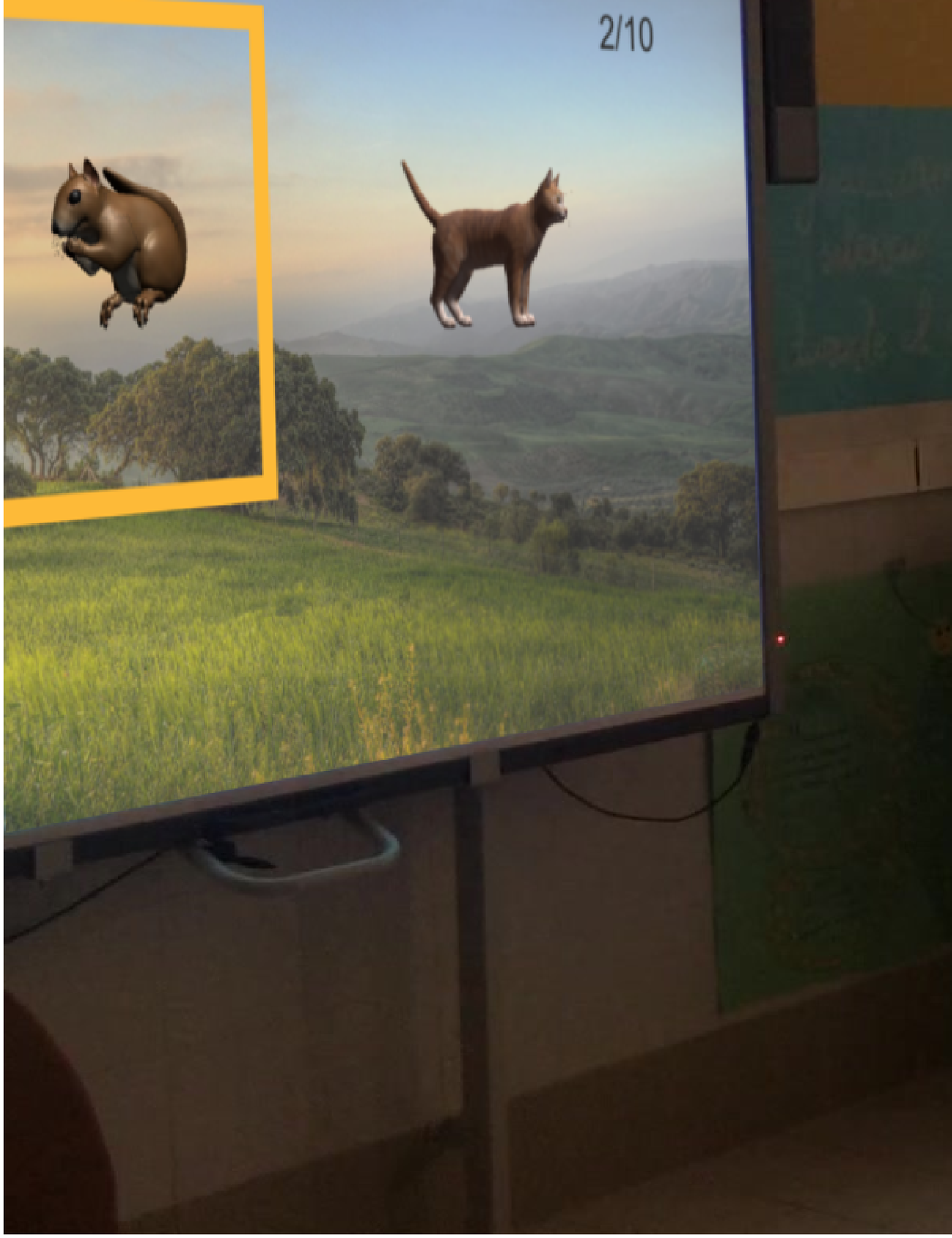}%
\label{f:hearingExperiment}}
$~~~~$
\subfloat[]{\includegraphics[width=2.2in]{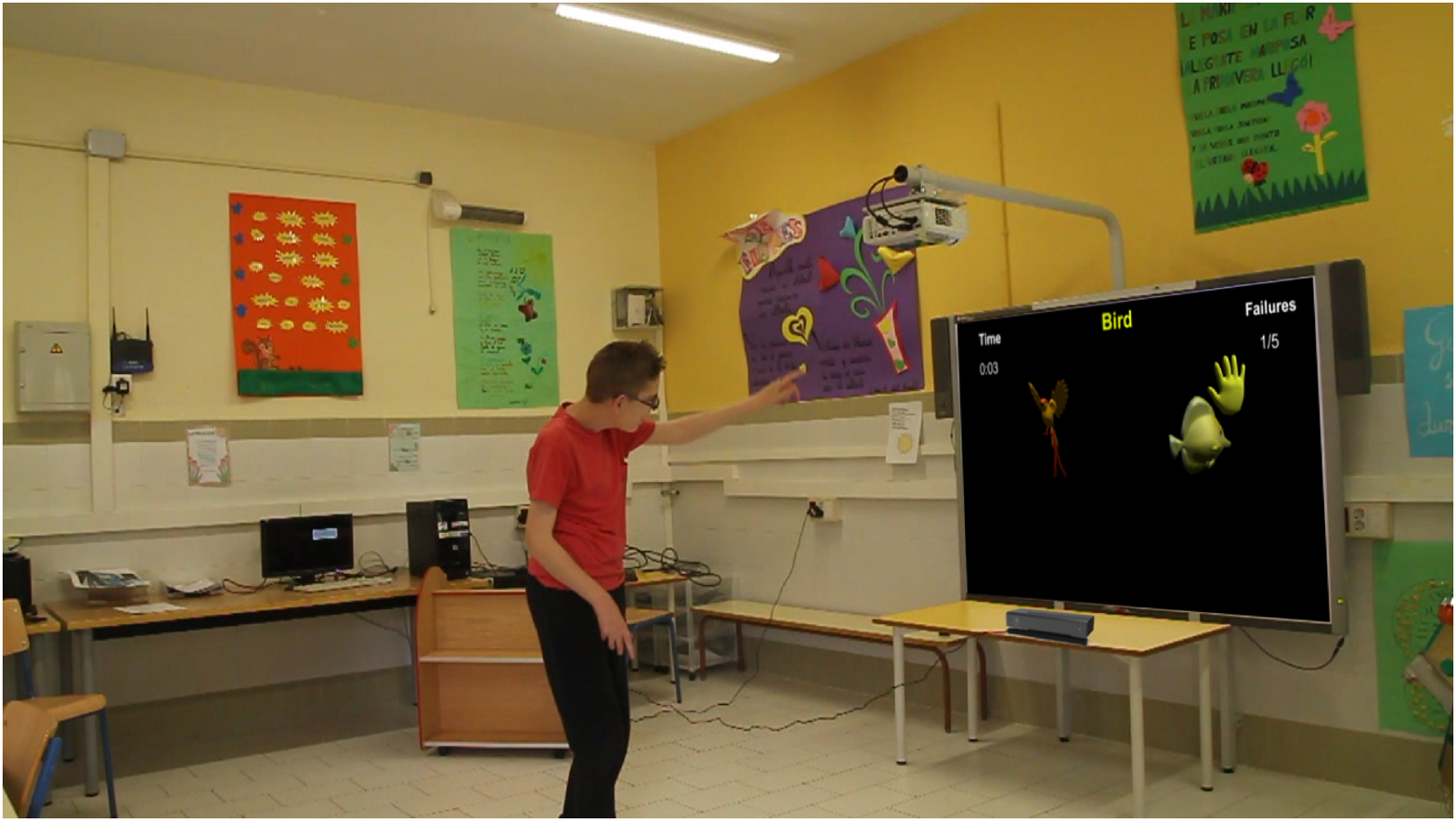}%
\label{f:visualExperiment}}
\hfill
\\
\centering
\subfloat[]{\includegraphics[width=2.2in]{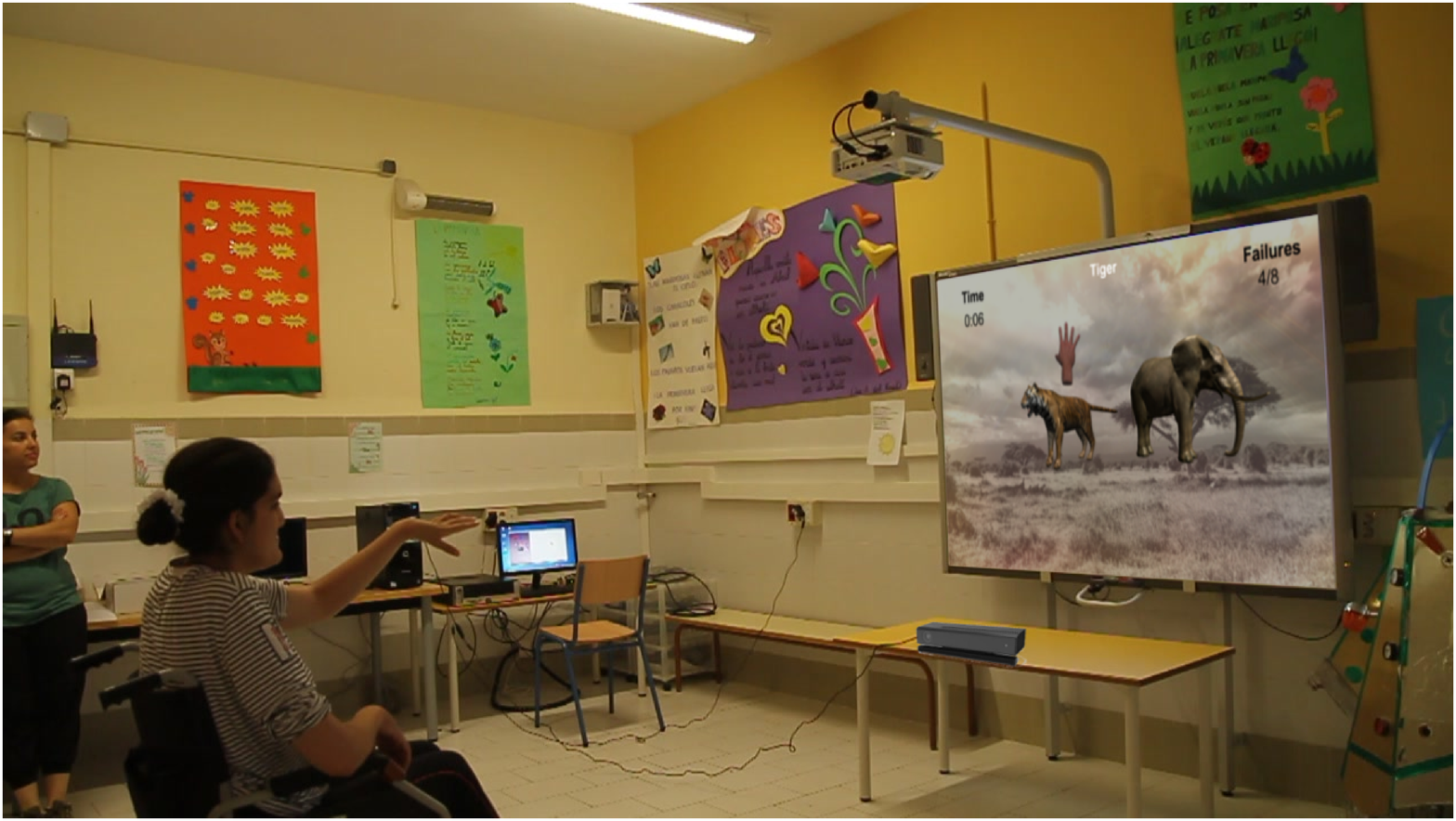}%
\label{f:physicalExperiment}}
$~~~~$
\subfloat[]{\includegraphics[width=2.2in]{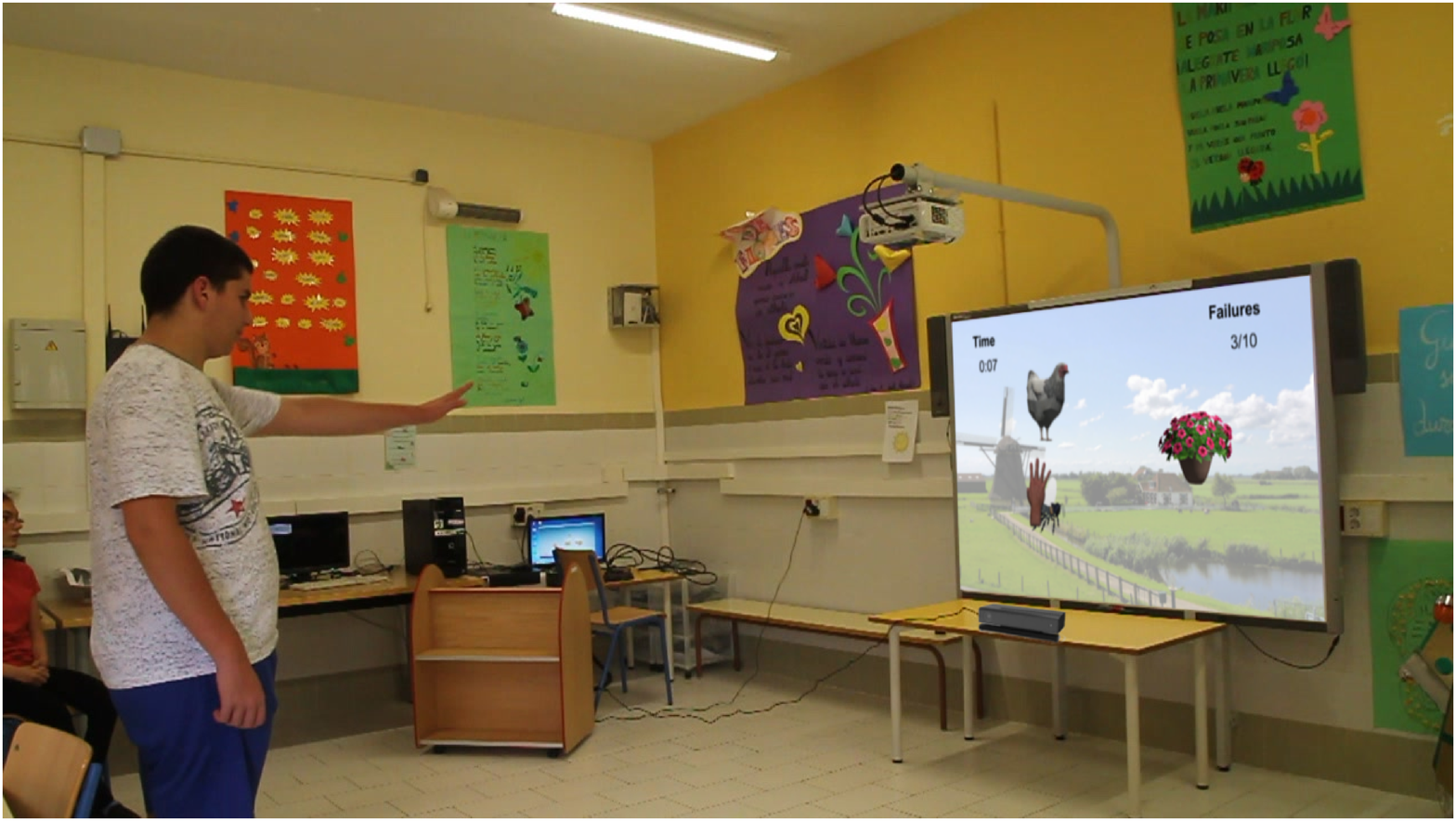}%
\label{f:autismExperiment}}
\caption{Experiments with the different cases.}
\label{screensExperiments}
\end{figure*}

Twelve students from the Special Educational Center Princesa Sofia participated in this evaluation. These students had different types of disability: three students with physical disability, three students with hearing impairment, three students with visual impairment and three students with autism. The evaluation had two iterations. In Figure \ref{screensExperiments}, experiments conducted in this evaluation process were shown.

In the first iteration, the same methodology for each student who participated in the evaluation was used. The process consisted of three sessions, each carried out on a different day. The test was carried out in a room where the students were used to doing very different types of activities because this room contains an interactive whiteboard, the projector, educational resources such as sheets, books and a couple of computers. Moreover, this room is the ideal place to interact with Kinect because the computer is connected to the whiteboard which is bigger than a conventional monitor and the room is very spacious. Thus, the learners can move freely and be as far away from the screen as is recommended for an optimal experience with this device. Individual students were in the room with their tutor, except in the case where several students had the same tutor. In this case, every student from the same tutor was in the room, but the prototype was only tried by one of the students while the rest of them were waiting. The first session lasted longer than other sessions in order to create the device-interaction model for each participant. The same activity was carried out in each session. There was no time limit to complete the activity but the activity was repeated ten times. Data collection was automatic because the system was connected to a database and every learner has a profile in the system. The data collected in this iteration were the time and the number of errors.

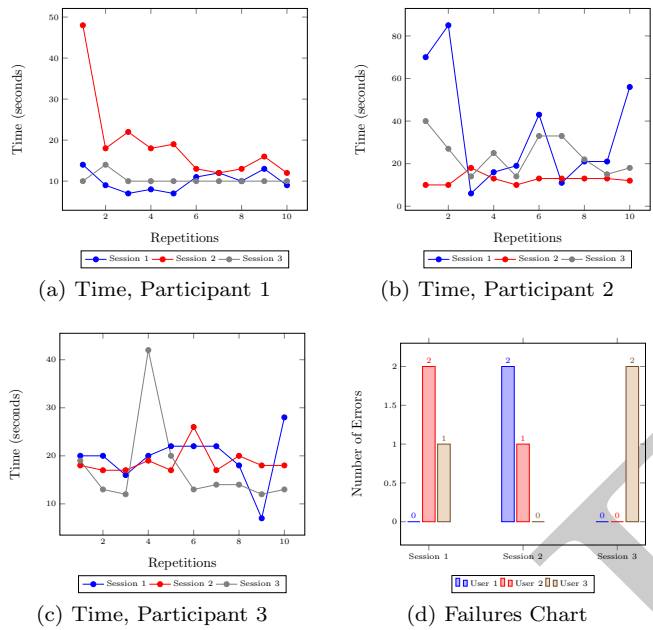
\begin{figure} 
\centering
{\fontsize{6.5}{7}\selectfont
\subfloat[Time, Participant 1]{
\begin{tikzpicture} [scale=0.47]
	\begin{axis}[
		xlabel=Repetitions,ylabel=Time (seconds), legend style={at={(0.5,-0.2)},
      anchor=north,legend columns=-1}, label style={font=\large}]

	\addplot[color=blue,mark=*] coordinates {
		(1,14)
		(2,9)
		(3,7)
		(4,8)
		(5,7)
		(6,11)
		(7,12)
		(8,10)
		(9,13)
		(10,9)
	};
	\addplot[color=red,mark=*] coordinates {
		(1,48)
		(2,18)
		(3,22)
		(4,18)
		(5,19)
		(6,13)
		(7,12)
		(8,13)
		(9,16)
		(10,12)
	};
	\addplot[color=gray,mark=*] coordinates {
		(1,10)
		(2,14)
		(3,10)
		(4,10)
		(5,10)
		(6,10)
		(7,10)
		(8,10)
		(9,10)
		(10,10)
	};
	\legend{Session 1, Session 2, Session 3}
	\end{axis}%
\end{tikzpicture}%
}
$~~~~$
\subfloat[Time, Participant 2]{
\begin{tikzpicture} [scale=0.47]
	\begin{axis}[
		xlabel=Repetitions,ylabel=Time (seconds), legend style={at={(0.5,-0.2)},
      anchor=north,legend columns=-1}, label style={font=\large}]

	\addplot[color=blue,mark=*] coordinates {
		(1,70)
		(2,85)
		(3,6)
		(4,16)
		(5,19)
		(6,43)
		(7,11)
		(8,21)
		(9,21)
		(10,56)
	};
	\addplot[color=red,mark=*] coordinates {
		(1,10)
		(2,10)
		(3,18)
		(4,13)
		(5,10)
		(6,13)
		(7,13)
		(8,13)
		(9,13)
		(10,12)
	};
	\addplot[color=gray,mark=*] coordinates {
		(1,40)
		(2,27)
		(3,14)
		(4,25)
		(5,14)
		(6,33)
		(7,33)
		(8,22)
		(9,15)
		(10,18)
	};
	\legend{Session 1, Session 2, Session 3}
	\end{axis}%
\end{tikzpicture}%
}

\subfloat[Time, Participant 3]{
\begin{tikzpicture}[scale=0.47]
	\begin{axis}[
		xlabel=Repetitions,ylabel=Time (seconds), legend style={at={(0.5,-0.2)},
      anchor=north,legend columns=-1}, label style={font=\large}]

	\addplot[color=blue,mark=*] coordinates {
		(1,20)
		(2,20)
		(3,16)
		(4,20)
		(5,22)
		(6,22)
		(7,22)
		(8,18)
		(9,7)
		(10,28)
	};
	\addplot[color=red,mark=*] coordinates {
		(1,18)
		(2,17)
		(3,17)
		(4,19)
		(5,17)
		(6,26)
		(7,17)
		(8,20)
		(9,18)
		(10,18)
	};
	\addplot[color=gray,mark=*] coordinates {
		(1,19)
		(2,13)
		(3,12)
		(4,42)
		(5,20)
		(6,13)
		(7,14)
		(8,14)
		(9,12)
		(10,13)
	};
	\legend{Session 1, Session 2, Session 3}
	\end{axis}%
\end{tikzpicture}%
}
$~~~~$
\subfloat[Failures Chart]{
\begin{tikzpicture}[scale=0.47]
\begin{axis}[
    ybar,
    enlargelimits=0.15,
    legend style={at={(0.5,-0.15)},
      anchor=north,legend columns=-1},
    ylabel={Number of Errors},
    symbolic x coords={Session 1,Session 2,Session 3},
    xtick=data,
    nodes near coords,
    nodes near coords align={vertical},
    label style={font=\large}
    ]
\addplot coordinates {(Session 1,0) (Session 2,2) (Session 3,0)};
\addplot coordinates {(Session 1,2) (Session 2,1) (Session 3,0)};
\addplot coordinates {(Session 1,1) (Session 2,0) (Session 3,2)};
\legend{User 1,User 2,User 3}
\end{axis}
\end{tikzpicture}
}
\caption{Participants with autism Charts in the $1^{st}$ iteration.}
\label{autismResults1st}
}
\end{figure}

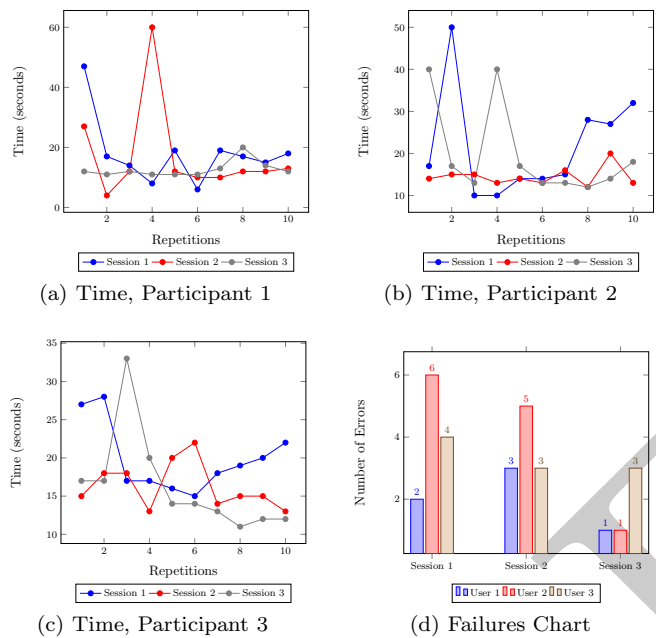
\begin{figure}[!h]
\centering
\subfloat[Time, Participant 1]{
\begin{tikzpicture}[scale=0.47]
	\begin{axis}[
		xlabel=Repetitions,ylabel=Time (seconds), legend style={at={(0.5,-0.2)},
      anchor=north,legend columns=-1}, label style={font=\large}]

	\addplot[color=blue,mark=*] coordinates {
		(1,47)
		(2,17)
		(3,14)
		(4,8)
		(5,19)
		(6,6)
		(7,19)
		(8,17)
		(9,15)
		(10,18)
	};
	\addplot[color=red,mark=*] coordinates {
		(1,27)
		(2,4)
		(3,12)
		(4,60)
		(5,12)
		(6,10)
		(7,10)
		(8,12)
		(9,12)
		(10,13)
	};
	\addplot[color=gray,mark=*] coordinates {
		(1,12)
		(2,11)
		(3,12)
		(4,11)
		(5,11)
		(6,11)
		(7,13)
		(8,20)
		(9,14)
		(10,12)
	};
	\legend{Session 1, Session 2, Session 3}
	\end{axis}%
\end{tikzpicture}%
}
$~~~~$
\subfloat[Time, Participant 2]{
\begin{tikzpicture}[scale=0.47]
	\begin{axis}[
		xlabel=Repetitions,ylabel=Time (seconds), legend style={at={(0.5,-0.2)},
      anchor=north,legend columns=-1}, label style={font=\large}]

	\addplot[color=blue,mark=*] coordinates {
		(1,17)
		(2,50)
		(3,10)
		(4,10)
		(5,14)
		(6,14)
		(7,15)
		(8,28)
		(9,27)
		(10,32)
	};
	\addplot[color=red,mark=*] coordinates {
		(1,14)
		(2,15)
		(3,15)
		(4,13)
		(5,14)
		(6,13)
		(7,16)
		(8,12)
		(9,20)
		(10,13)
	};
	\addplot[color=gray,mark=*] coordinates {
		(1,40)
		(2,17)
		(3,13)
		(4,40)
		(5,17)
		(6,13)
		(7,13)
		(8,12)
		(9,14)
		(10,18)
	};
	\legend{Session 1, Session 2, Session 3}
	\end{axis}%
\end{tikzpicture}%
}

\subfloat[Time, Participant 3]{
\begin{tikzpicture}[scale=0.47]
	\begin{axis}[
		xlabel=Repetitions,ylabel=Time (seconds), legend style={at={(0.5,-0.2)},
      anchor=north,legend columns=-1}, label style={font=\large}]

	\addplot[color=blue,mark=*] coordinates {
		(1,27)
		(2,28)
		(3,17)
		(4,17)
		(5,16)
		(6,15)
		(7,18)
		(8,19)
		(9,20)
		(10,22)
	};
	\addplot[color=red,mark=*] coordinates {
		(1,15)
		(2,18)
		(3,18)
		(4,13)
		(5,20)
		(6,22)
		(7,14)
		(8,15)
		(9,15)
		(10,13)
	};
	\addplot[color=gray,mark=*] coordinates {
		(1,17)
		(2,17)
		(3,33)
		(4,20)
		(5,14)
		(6,14)
		(7,13)
		(8,11)
		(9,12)
		(10,12)
	};
	\legend{Session 1, Session 2, Session 3}
	\end{axis}%
\end{tikzpicture}%
}
$~~~~$
\subfloat[Failures Chart]{
\begin{tikzpicture}[scale=0.47]
\begin{axis}[
    ybar,
    enlargelimits=0.15,
    legend style={at={(0.5,-0.15)},
      anchor=north,legend columns=-1},
    ylabel={Number of Errors},
    symbolic x coords={Session 1,Session 2,Session 3},
    xtick=data,
    nodes near coords,
    nodes near coords align={vertical},
    label style={font=\large}
    ]
\addplot coordinates {(Session 1,2) (Session 2,3) (Session 3,1)};
\addplot coordinates {(Session 1,6) (Session 2,5) (Session 3,1)};
\addplot coordinates {(Session 1,4) (Session 2,3) (Session 3,3)};
\legend{User 1,User 2,User 3}
\end{axis}
\end{tikzpicture}
}
\caption{Hearing Impairment Participants Results in the $1^{st}$ iteration.}
\label{hearingResults1st}
\end{figure}

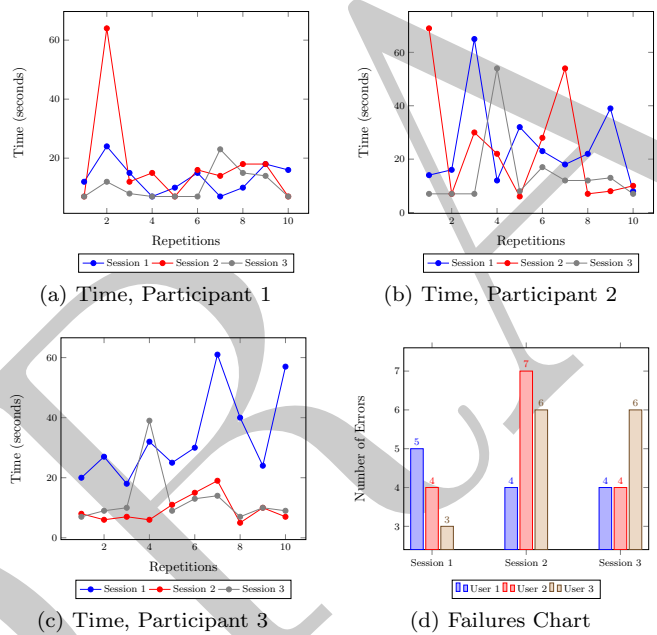
\begin{figure}[!h]
\centering
\subfloat[Time, Participant 1]{
\begin{tikzpicture}[scale=0.47]
	\begin{axis}[
		xlabel=Repetitions,ylabel=Time (seconds), legend style={at={(0.5,-0.2)},
      anchor=north,legend columns=-1}, label style={font=\large}]

	\addplot[color=blue,mark=*] coordinates {
		(1,12)
		(2,24)
		(3,15)
		(4,7)
		(5,10)
		(6,15)
		(7,7)
		(8,10)
		(9,18)
		(10,16)
	};
	\addplot[color=red,mark=*] coordinates {
		(1,7)
		(2,64)
		(3,12)
		(4,15)
		(5,7)
		(6,16)
		(7,14)
		(8,18)
		(9,18)
		(10,7)
	};
	\addplot[color=gray,mark=*] coordinates {
		(1,7)
		(2,12)
		(3,8)
		(4,7)
		(5,7)
		(6,7)
		(7,23)
		(8,15)
		(9,14)
		(10,7)
	};
	\legend{Session 1, Session 2, Session 3}
	\end{axis}%
\end{tikzpicture}%
}
$~~~~$
\subfloat[Time, Participant 2]{
\begin{tikzpicture}[scale=0.47]
	\begin{axis}[
		xlabel=Repetitions,ylabel=Time (seconds), legend style={at={(0.5,-0.2)},
      anchor=north,legend columns=-1}, label style={font=\large}]

	\addplot[color=blue,mark=*] coordinates {
		(1,14)
		(2,16)
		(3,65)
		(4,12)
		(5,32)
		(6,23)
		(7,18)
		(8,22)
		(9,39)
		(10,8)
	};
	\addplot[color=red,mark=*] coordinates {
		(1,69)
		(2,7)
		(3,30)
		(4,22)
		(5,6)
		(6,28)
		(7,54)
		(8,7)
		(9,8)
		(10,10)
	};
	\addplot[color=gray,mark=*] coordinates {
		(1,7)
		(2,7)
		(3,7)
		(4,54)
		(5,8)
		(6,17)
		(7,12)
		(8,12)
		(9,13)
		(10,7)
	};
	\legend{Session 1, Session 2, Session 3}
	\end{axis}%
	\label{physicalResults1st_part2}
\end{tikzpicture}%
}

\subfloat[Time, Participant 3]{
\begin{tikzpicture}[scale=0.47]
	\begin{axis}[
		xlabel=Repetitions,ylabel=Time (seconds), legend style={at={(0.5,-0.2)},
      anchor=north,legend columns=-1}, label style={font=\large}]

	\addplot[color=blue,mark=*] coordinates {
		(1,20)
		(2,27)
		(3,18)
		(4,32)
		(5,25)
		(6,30)
		(7,61)
		(8,40)
		(9,24)
		(10,57)
	};
	\addplot[color=red,mark=*] coordinates {
		(1,8)
		(2,6)
		(3,7)
		(4,6)
		(5,11)
		(6,15)
		(7,19)
		(8,5)
		(9,10)
		(10,7)
	};
	\addplot[color=gray,mark=*] coordinates {
		(1,7)
		(2,9)
		(3,10)
		(4,39)
		(5,9)
		(6,13)
		(7,14)
		(8,7)
		(9,10)
		(10,9)
	};
	\legend{Session 1, Session 2, Session 3}
	\end{axis}%
\end{tikzpicture}%
}
$~~~~$
\subfloat[Failures Chart]{
\begin{tikzpicture}[scale=0.47]
\begin{axis}[
    ybar,
    enlargelimits=0.15,
    legend style={at={(0.5,-0.15)},
      anchor=north,legend columns=-1},
    ylabel={Number of Errors},
    symbolic x coords={Session 1,Session 2,Session 3},
    xtick=data,
    nodes near coords,
    nodes near coords align={vertical},
    label style={font=\large}
    ]
\addplot coordinates {(Session 1,5) (Session 2,4) (Session 3,4)};
\addplot coordinates {(Session 1,4) (Session 2,7) (Session 3,4)};
\addplot coordinates {(Session 1,3) (Session 2,6) (Session 3,6)};
\legend{User 1,User 2,User 3}
\end{axis}
\end{tikzpicture}
}
\caption{Physical Impairment Participants Results in the $1^{st}$ iteration.}
\label{physicalResults1st}
\end{figure}

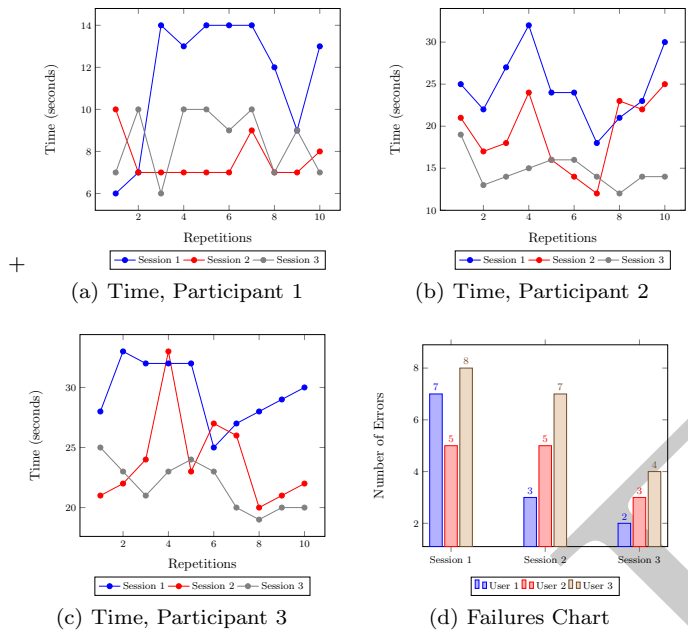
\begin{figure} [!h]
\centering+
\subfloat[Time, Participant 1]{
\begin{tikzpicture}[scale=0.47]
	\begin{axis}[
		xlabel=Repetitions,ylabel=Time (seconds), legend style={at={(0.5,-0.2)},
      anchor=north,legend columns=-1}, label style={font=\large}]

	\addplot[color=blue,mark=*] coordinates {
		(1,6)
		(2,7)
		(3,14)
		(4,13)
		(5,14)
		(6,14)
		(7,14)
		(8,12)
		(9,9)
		(10,13)
	};
	\addplot[color=red,mark=*] coordinates {
		(1,10)
		(2,7)
		(3,7)
		(4,7)
		(5,7)
		(6,7)
		(7,9)
		(8,7)
		(9,7)
		(10,8)
	};
	\addplot[color=gray,mark=*] coordinates {
		(1,7)
		(2,10)
		(3,6)
		(4,10)
		(5,10)
		(6,9)
		(7,10)
		(8,7)
		(9,9)
		(10,7)
	};
	\legend{Session 1, Session 2, Session 3}
	\end{axis}%
\end{tikzpicture}%
}
$~~~~$
\subfloat[Time, Participant 2]{
\begin{tikzpicture}[scale=0.47]
	\begin{axis}[
		xlabel=Repetitions,ylabel=Time (seconds), legend style={at={(0.5,-0.2)},
      anchor=north,legend columns=-1}, label style={font=\large}]

	\addplot[color=blue,mark=*] coordinates {
		(1,25)
		(2,22)
		(3,27)
		(4,32)
		(5,24)
		(6,24)
		(7,18)
		(8,21)
		(9,23)
		(10,30)
	};
	\addplot[color=red,mark=*] coordinates {
		(1,21)
		(2,17)
		(3,18)
		(4,24)
		(5,16)
		(6,14)
		(7,12)
		(8,23)
		(9,22)
		(10,25)
	};
	\addplot[color=gray,mark=*] coordinates {
		(1,19)
		(2,13)
		(3,14)
		(4,15)
		(5,16)
		(6,16)
		(7,14)
		(8,12)
		(9,14)
		(10,14)
	};
	\legend{Session 1, Session 2, Session 3}
	\end{axis}%
\end{tikzpicture}%
}

\subfloat[Time, Participant 3]{
\begin{tikzpicture}[scale=0.47]
	\begin{axis}[
		xlabel=Repetitions,ylabel=Time (seconds), legend style={at={(0.5,-0.2)},
      anchor=north,legend columns=-1}, label style={font=\large}]

	\addplot[color=blue,mark=*] coordinates {
		(1,28)
		(2,33)
		(3,32)
		(4,32)
		(5,32)
		(6,25)
		(7,27)
		(8,28)
		(9,29)
		(10,30)
	};
	\addplot[color=red,mark=*] coordinates {
		(1,21)
		(2,22)
		(3,24)
		(4,33)
		(5,23)
		(6,27)
		(7,26)
		(8,20)
		(9,21)
		(10,22)
	};
	\addplot[color=gray,mark=*] coordinates {
		(1,25)
		(2,23)
		(3,21)
		(4,23)
		(5,24)
		(6,23)
		(7,20)
		(8,19)
		(9,20)
		(10,20)
	};
	\legend{Session 1, Session 2, Session 3}
	\end{axis}%
\end{tikzpicture}%
}
$~~~~$
\subfloat[Failures Chart]{
\begin{tikzpicture}[scale=0.47]
\begin{axis}[
    ybar,
    enlargelimits=0.15,
    legend style={at={(0.5,-0.15)},
      anchor=north,legend columns=-1},
    ylabel={Number of Errors},
    symbolic x coords={Session 1,Session 2,Session 3},
    xtick=data,
    nodes near coords,
    nodes near coords align={vertical},
    label style={font=\large}
    ]
\addplot coordinates {(Session 1,7) (Session 2,3) (Session 3,2)};
\addplot coordinates {(Session 1,5) (Session 2,5) (Session 3,3)};
\addplot coordinates {(Session 1,8) (Session 2,7) (Session 3,4)};
\legend{User 1,User 2,User 3}
\end{axis}
\end{tikzpicture}
}
\caption{Visual Impairment Participants Results $1^{st}$ iteration.}
\label{visualResults1st}
\end{figure}

The next charts show the data from the user, grouped according to the different types of disability. Figure \ref{autismResults1st} portrays the data related to users with autism; Figures \ref{hearingResults1st} shows the data related to users with severe hearing loss. The data pertaining to users with physical disability is shown in Figure \ref{physicalResults1st} and for users with visual impairment, the data can be seen in Figure \ref{visualResults1st}.


In the first iteration, a reduction in time and the number of errors made by the users can be seen in the different user groups that have participated in the experiment. These results are encouraging because the reduction in the errors means that the activity was carried out conscientiously and the improvement in time is not a consequence of random choices made  to finish the exercise early.

The users felt comfortable and more confident with respect to the interaction the more they participated in the sessions. An important factor in this iteration was that users from the different groups could complete the activity since each user had very different characteristics, for example, wheelchair-bound. In spite of the promising results during these sessions, it is necessary to highlight the following aspects:

\begin{itemize}
\item In the tests carried out with the students with severe hearing loss, it is obvious that although the text instructions are shown, there are cases where the tutor has to indicate how to do the exercise. Therefore, it is recommendable that these instructions are shown in a graphical way, even with a cartoon.
\item In the case where the user has a physical disability, it is worth focusing on the irregularity of the second participant (see Figure \ref{physicalResults1st_part2}) in the chart. This situation is due to the student losing interest quickly and as a consequence, some repetitions were completed in a reasonable time while in other repetitions, the time taken increased considerably.
\item The learners with autism did not find it particularly difficult to understand and adapt to the exercise as can be seen in Figure \ref{autismResults1st}, where participants 1 and 2 completed the last session without any errors.
\end{itemize}

In the second iteration, the evaluation room had been changed and the tests were carried out in a specific room which was appointed by the head of the center to do the activities with Kinect. This decision was made because in the time charts of the previous iteration there is a very large gap between the values in the same session of every group. These gaps were caused by the different elements (toys, posters and so on) that were in the room and were distracting the students. Therefore, this new room was painted white and only contained the essential equipment to do the exercises with this device: a desktop computer with some speakers, a 42-inch monitor and the Kinect device. As in the previous iteration, all the students were in the room with their tutors. The only difference was that the students had to wait outside for their turn. Thus, the sessions were all conducted individually. In contrast to the previous iteration, two activities were performed in each session: the concept association activity (changing the topic between animals and transport) and the laterality activity.

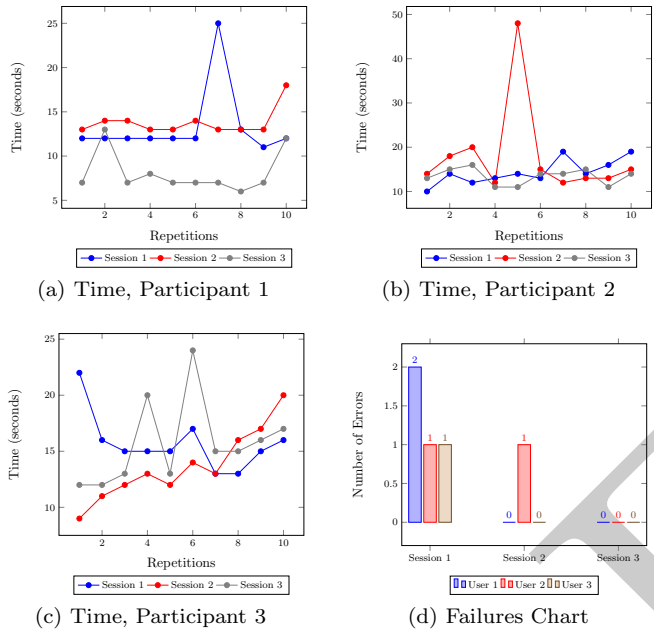
\begin{figure}
\centering
\subfloat[Time, Participant 1]{
\begin{tikzpicture}[scale=0.47]
	\begin{axis}[
		xlabel=Repetitions,ylabel=Time (seconds), legend style={at={(0.5,-0.2)},
      anchor=north,legend columns=-1}, label style={font=\large}]

	\addplot[color=blue,mark=*] coordinates {
		(1,12)
		(2,12)
		(3,12)
		(4,12)
		(5,12)
		(6,12)
		(7,25)
		(8,13)
		(9,11)
		(10,12)
	};
	\addplot[color=red,mark=*] coordinates {
		(1,13)
		(2,14)
		(3,14)
		(4,13)
		(5,13)
		(6,14)
		(7,13)
		(8,13)
		(9,13)
		(10,18)
	};
	\addplot[color=gray,mark=*] coordinates {
		(1,7)
		(2,13)
		(3,7)
		(4,8)
		(5,7)
		(6,7)
		(7,7)
		(8,6)
		(9,7)
		(10,12)
	};
	\legend{Session 1, Session 2, Session 3}
	\end{axis}%
\end{tikzpicture}%
}
$~~~~$
\subfloat[Time, Participant 2]{
\begin{tikzpicture}[scale=0.47]
	\begin{axis}[
		xlabel=Repetitions,ylabel=Time (seconds), legend style={at={(0.5,-0.2)},
      anchor=north,legend columns=-1}, label style={font=\large}]

	\addplot[color=blue,mark=*] coordinates {
		(1,10)
		(2,14)
		(3,12)
		(4,13)
		(5,14)
		(6,13)
		(7,19)
		(8,14)
		(9,16)
		(10,19)
	};
	\addplot[color=red,mark=*] coordinates {
		(1,14)
		(2,18)
		(3,20)
		(4,12)
		(5,48)
		(6,15)
		(7,12)
		(8,13)
		(9,13)
		(10,15)
	};
	\addplot[color=gray,mark=*] coordinates {
		(1,13)
		(2,15)
		(3,16)
		(4,11)
		(5,11)
		(6,14)
		(7,14)
		(8,15)
		(9,11)
		(10,14)
	};
	\legend{Session 1, Session 2, Session 3}
	\end{axis}%
\end{tikzpicture}%
}

\subfloat[Time, Participant 3]{
\begin{tikzpicture}[scale=0.47]
	\begin{axis}[
		xlabel=Repetitions,ylabel=Time (seconds), legend style={at={(0.5,-0.2)},
      anchor=north,legend columns=-1}, label style={font=\large}]

	\addplot[color=blue,mark=*] coordinates {
		(1,22)
		(2,16)
		(3,15)
		(4,15)
		(5,15)
		(6,17)
		(7,13)
		(8,13)
		(9,15)
		(10,16)
	};
	\addplot[color=red,mark=*] coordinates {
		(1,9)
		(2,11)
		(3,12)
		(4,13)
		(5,12)
		(6,14)
		(7,13)
		(8,16)
		(9,17)
		(10,20)
	};
	\addplot[color=gray,mark=*] coordinates {
		(1,12)
		(2,12)
		(3,13)
		(4,20)
		(5,13)
		(6,24)
		(7,15)
		(8,15)
		(9,16)
		(10,17)
	};
	\legend{Session 1, Session 2, Session 3}
	\end{axis}%
\end{tikzpicture}%
}
$~~~~$
\subfloat[Failures Chart]{
\begin{tikzpicture}[scale=0.47]
\begin{axis}[
    ybar,
    enlargelimits=0.15,
    legend style={at={(0.5,-0.15)},
      anchor=north,legend columns=-1},
    ylabel={Number of Errors},
    symbolic x coords={Session 1,Session 2,Session 3},
    xtick=data,
    nodes near coords,
    nodes near coords align={vertical},
    label style={font=\large}
    ]
\addplot coordinates {(Session 1,2) (Session 2,0) (Session 3,0)};
\addplot coordinates {(Session 1,1) (Session 2,1) (Session 3,0)};
\addplot coordinates {(Session 1,1) (Session 2,0) (Session 3,0)};
\legend{User 1,User 2,User 3}
\end{axis}
\label{autismResults2nd_errors}
\end{tikzpicture}
}
\caption{Participants with Autism Results in the $2^{nd}$ iteration.}
\label{autismResults2nd}
\end{figure}

\begin{figure} 
\centering
\subfloat[Time, Participant 1]{
\begin{tikzpicture}[scale=0.47]
	\begin{axis}[
		xlabel=Repetitions,ylabel=Time (seconds), legend style={at={(0.5,-0.2)},
      anchor=north,legend columns=-1}, label style={font=\large}]

	\addplot[color=blue,mark=*] coordinates {
		(1,11)
		(2,20)
		(3,31)
		(4,22)
		(5,17)
		(6,16)
		(7,13)
		(8,15)
		(9,12)
		(10,39)
	};
	\addplot[color=red,mark=*] coordinates {
		(1,12)
		(2,12)
		(3,13)
		(4,12)
		(5,13)
		(6,15)
		(7,13)
		(8,16)
		(9,13)
		(10,12)
	};
	\addplot[color=gray,mark=*] coordinates {
		(1,12)
		(2,20)
		(3,13)
		(4,10)
		(5,17)
		(6,19)
		(7,10)
		(8,10)
		(9,10)
		(10,10)
	};
	\legend{Session 1, Session 2, Session 3}
	\end{axis}%
\end{tikzpicture}%
}
$~~~~$
\subfloat[Time, Participant 2]{
\begin{tikzpicture}[scale=0.47]
	\begin{axis}[
		xlabel=Repetitions,ylabel=Time (seconds), legend style={at={(0.5,-0.2)},
      anchor=north,legend columns=-1}, label style={font=\large}]

	\addplot[color=blue,mark=*] coordinates {
		(1,17)
		(2,41)
		(3,12)
		(4,19)
		(5,12)
		(6,14)
		(7,16)
		(8,12)
		(9,13)
		(10,13)
	};
	\addplot[color=red,mark=*] coordinates {
		(1,21)
		(2,27)
		(3,17)
		(4,13)
		(5,13)
		(6,18)
		(7,19)
		(8,12)
		(9,16)
		(10,18)
	};
	\addplot[color=gray,mark=*] coordinates {
		(1,12)
		(2,15)
		(3,14)
		(4,11)
		(5,15)
		(6,16)
		(7,15)
		(8,14)
		(9,10)
		(10,15)
	};
	\legend{Session 1, Session 2, Session 3}
	\end{axis}%
\end{tikzpicture}%
}

\subfloat[Time, Participant 3]{
\begin{tikzpicture}[scale=0.47]
	\begin{axis}[
		xlabel=Repetitions,ylabel=Time (seconds), legend style={at={(0.5,-0.2)},
      anchor=north,legend columns=-1}, label style={font=\large}]

	\addplot[color=blue,mark=*] coordinates {
		(1,23)
		(2,19)
		(3,21)
		(4,18)
		(5,14)
		(6,15)
		(7,17)
		(8,25)
		(9,15)
		(10,14)
	};
	\addplot[color=red,mark=*] coordinates {
		(1,27)
		(2,26)
		(3,23)
		(4,24)
		(5,28)
		(6,26)
		(7,22)
		(8,24)
		(9,24)
		(10,23)
	};
	\addplot[color=gray,mark=*] coordinates {
		(1,18)
		(2,15)
		(3,15)
		(4,16)
		(5,15)
		(6,14)
		(7,20)
		(8,13)
		(9,13)
		(10,17)
	};
	\legend{Session 1, Session 2, Session 3}
	\end{axis}%
\end{tikzpicture}%
}
$~~~~$
\subfloat[Failures Chart]{
\begin{tikzpicture}[scale=0.47]
\begin{axis}[
    ybar,
    enlargelimits=0.15,
    legend style={at={(0.5,-0.15)},
      anchor=north,legend columns=-1},
    ylabel={Number of Errors},
    symbolic x coords={Session 1,Session 2,Session 3},
    xtick=data,
    nodes near coords,
    nodes near coords align={vertical},
    label style={font=\large}
    ]
\addplot coordinates {(Session 1,2) (Session 2,2) (Session 3,1)};
\addplot coordinates {(Session 1,4) (Session 2,7) (Session 3,5)};
\addplot coordinates {(Session 1,3) (Session 2,3) (Session 3,2)};
\legend{User 1,User 2,User 3}
\end{axis}
\label{hearingResults2nd_errors}
\end{tikzpicture}
}
\caption{Hearing Impairment Participants Results in the $2^{nd}$ iteration.}
\label{hearingResults2nd}
\end{figure}
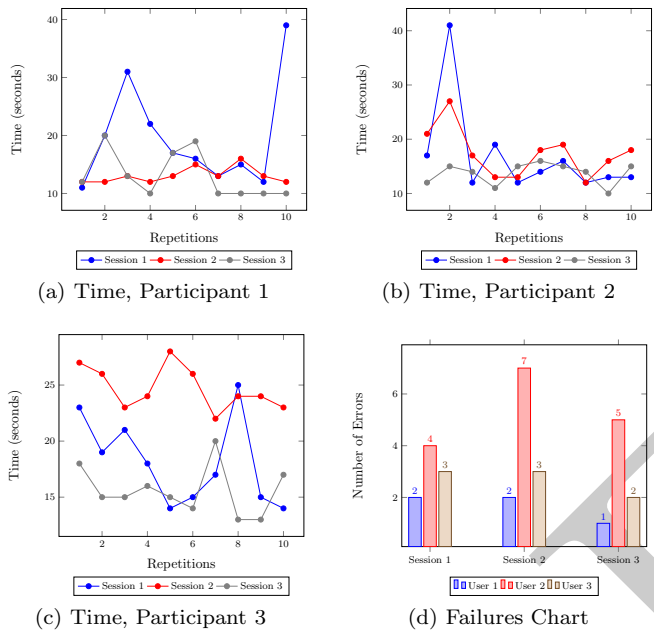

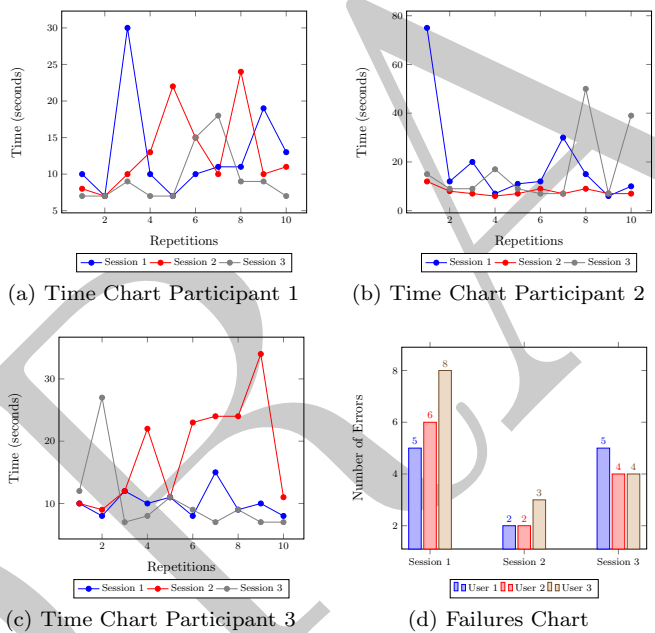
\begin{figure} 
\centering
\subfloat[Time Chart Participant 1]{
\begin{tikzpicture}[scale=0.47]
	\begin{axis}[
		xlabel=Repetitions,ylabel=Time (seconds), legend style={at={(0.5,-0.2)},
      anchor=north,legend columns=-1}, label style={font=\large}]

	\addplot[color=blue,mark=*] coordinates {
		(1,10)
		(2,7)
		(3,30)
		(4,10)
		(5,7)
		(6,10)
		(7,11)
		(8,11)
		(9,19)
		(10,13)
	};
	\addplot[color=red,mark=*] coordinates {
		(1,8)
		(2,7)
		(3,10)
		(4,13)
		(5,22)
		(6,15)
		(7,10)
		(8,24)
		(9,10)
		(10,11)
	};
	\addplot[color=gray,mark=*] coordinates {
		(1,7)
		(2,7)
		(3,9)
		(4,7)
		(5,7)
		(6,15)
		(7,18)
		(8,9)
		(9,9)
		(10,7)
	};
	\legend{Session 1, Session 2, Session 3}
	\end{axis}%
\end{tikzpicture}%
}
$~~~~$
\subfloat[Time Chart Participant 2]{
\begin{tikzpicture}[scale=0.47]
	\begin{axis}[
		xlabel=Repetitions,ylabel=Time (seconds), legend style={at={(0.5,-0.2)},
      anchor=north,legend columns=-1}, label style={font=\large}]

	\addplot[color=blue,mark=*] coordinates {
		(1,75)
		(2,12)
		(3,20)
		(4,7)
		(5,11)
		(6,12)
		(7,30)
		(8,15)
		(9,6)
		(10,10)
	};
	\addplot[color=red,mark=*] coordinates {
		(1,12)
		(2,8)
		(3,7)
		(4,6)
		(5,7)
		(6,9)
		(7,7)
		(8,9)
		(9,7)
		(10,7)
	};
	\addplot[color=gray,mark=*] coordinates {
		(1,15)
		(2,9)
		(3,9)
		(4,17)
		(5,9)
		(6,7)
		(7,7)
		(8,50)
		(9,7)
		(10,39)
	};
	\legend{Session 1, Session 2, Session 3}
	\end{axis}%
\end{tikzpicture}%
}

\subfloat[Time Chart Participant 3]{
\begin{tikzpicture}[scale=0.47]
	\begin{axis}[
		xlabel=Repetitions,ylabel=Time (seconds), legend style={at={(0.5,-0.2)},
      anchor=north,legend columns=-1}, label style={font=\large}]

	\addplot[color=blue,mark=*] coordinates {
		(1,10)
		(2,8)
		(3,12)
		(4,10)
		(5,11)
		(6,8)
		(7,15)
		(8,9)
		(9,10)
		(10,8)
	};
	\addplot[color=red,mark=*] coordinates {
		(1,10)
		(2,9)
		(3,12)
		(4,22)
		(5,11)
		(6,23)
		(7,24)
		(8,24)
		(9,34)
		(10,11)
	};
	\addplot[color=gray,mark=*] coordinates {
		(1,12)
		(2,27)
		(3,7)
		(4,8)
		(5,11)
		(6,9)
		(7,7)
		(8,9)
		(9,7)
		(10,7)
	};
	\legend{Session 1, Session 2, Session 3}
	\end{axis}%
\end{tikzpicture}%
}
$~~~~$
\subfloat[Failures Chart]{
\begin{tikzpicture}[scale=0.47]
\begin{axis}[
    ybar,
    enlargelimits=0.15,
    legend style={at={(0.5,-0.15)},
      anchor=north,legend columns=-1},
    ylabel={Number of Errors},
    symbolic x coords={Session 1,Session 2,Session 3},
    xtick=data,
    nodes near coords,
    nodes near coords align={vertical},
    label style={font=\large}
    ]
\addplot coordinates {(Session 1,5) (Session 2,2) (Session 3,5)};
\addplot coordinates {(Session 1,6) (Session 2,2) (Session 3,4)};
\addplot coordinates {(Session 1,8) (Session 2,3) (Session 3,4)};
\legend{User 1,User 2,User 3}
\end{axis}
\label{physicalResults2nd_errors}
\end{tikzpicture}
}
\caption{Physical Impairment Participants Results in the $2^{nd}$ iteration.}
\label{physicalResults2nd}
\end{figure}

\begin{figure} 
\centering
\subfloat[Time, Participant 1]{
\begin{tikzpicture}[scale=0.47]
	\begin{axis}[
		xlabel=Repetitions,ylabel=Time (seconds), legend style={at={(0.5,-0.2)},
      anchor=north,legend columns=-1}, label style={font=\large}]

	\addplot[color=blue,mark=*] coordinates {
		(1,13)
		(2,10)
		(3,7)
		(4,10)
		(5,15)
		(6,7)
		(7,7)
		(8,8)
		(9,26)
		(10,14)
	};
	\addplot[color=red,mark=*] coordinates {
		(1,12)
		(2,7)
		(3,7)
		(4,12)
		(5,9)
		(6,7)
		(7,10)
		(8,8)
		(9,8)
		(10,7)
	};
	\addplot[color=gray,mark=*] coordinates {
		(1,7)
		(2,10)
		(3,13)
		(4,18)
		(5,9)
		(6,7)
		(7,9)
		(8,7)
		(9,21)
		(10,10)
	};
	\legend{Session 1, Session 2, Session 3}
	\end{axis}%
\end{tikzpicture}%
}
$~~~~$
\subfloat[Time, Participant 2]{
\begin{tikzpicture}[scale=0.47]
	\begin{axis}[
		xlabel=Repetitions,ylabel=Time (seconds), legend style={at={(0.5,-0.2)},
      anchor=north,legend columns=-1}, label style={font=\large}]

	\addplot[color=blue,mark=*] coordinates {
		(1,9)
		(2,12)
		(3,10)
		(4,7)
		(5,9)
		(6,34)
		(7,9)
		(8,26)
		(9,12)
		(10,8)
	};
	\addplot[color=red,mark=*] coordinates {
		(1,17)
		(2,9)
		(3,12)
		(4,7)
		(5,9)
		(6,8)
		(7,8)
		(8,10)
		(9,7)
		(10,8)
	};
	\addplot[color=gray,mark=*] coordinates {
		(1,7)
		(2,7)
		(3,7)
		(4,7)
		(5,7)
		(6,7)
		(7,9)
		(8,19)
		(9,7)
		(10,16)
	};
	\legend{Session 1, Session 2, Session 3}
	\end{axis}%
\end{tikzpicture}%
}

\subfloat[Time, Participant 3]{
\begin{tikzpicture}[scale=0.47]
	\begin{axis}[
		xlabel=Repetitions,ylabel=Time (seconds), legend style={at={(0.5,-0.2)},
      anchor=north,legend columns=-1}, label style={font=\large}]

	\addplot[color=blue,mark=*] coordinates {
		(1,18)
		(2,20)
		(3,22)
		(4,15)
		(5,23)
		(6,15)
		(7,17)
		(8,17)
		(9,21)
		(10,20)
	};
	\addplot[color=red,mark=*] coordinates {
		(1,11)
		(2,16)
		(3,14)
		(4,23)
		(5,13)
		(6,17)
		(7,20)
		(8,14)
		(9,11)
		(10,11)
	};
	\addplot[color=gray,mark=*] coordinates {
		(1,14)
		(2,15)
		(3,11)
		(4,13)
		(5,15)
		(6,15)
		(7,10)
		(8,9)
		(9,12)
		(10,10)
	};
	\legend{Session 1, Session 2, Session 3}
	\end{axis}%
\end{tikzpicture}%
}
$~~~~$
\subfloat[Failures Chart]{
\begin{tikzpicture}[scale=0.47]
\begin{axis}[
    ybar,
    enlargelimits=0.15,
    legend style={at={(0.5,-0.15)},
      anchor=north,legend columns=-1},
    ylabel={Number of Errors},
    symbolic x coords={Session 1,Session 2,Session 3},
    xtick=data,
    nodes near coords,
    nodes near coords align={vertical},
    label style={font=\large}
    ]
\addplot coordinates {(Session 1,2) (Session 2,2) (Session 3,1)};
\addplot coordinates {(Session 1,3) (Session 2,4) (Session 3,4)};
\addplot coordinates {(Session 1,5) (Session 2,4) (Session 3,2)};
\legend{User 1,User 2,User 3}
\end{axis}
\label{visualResults2nd_errors}
\end{tikzpicture}
}
\caption{Visual Impairment Participants Results in the - $2^{nd}$ iteration.}
\label{visualResults2nd}
\end{figure}
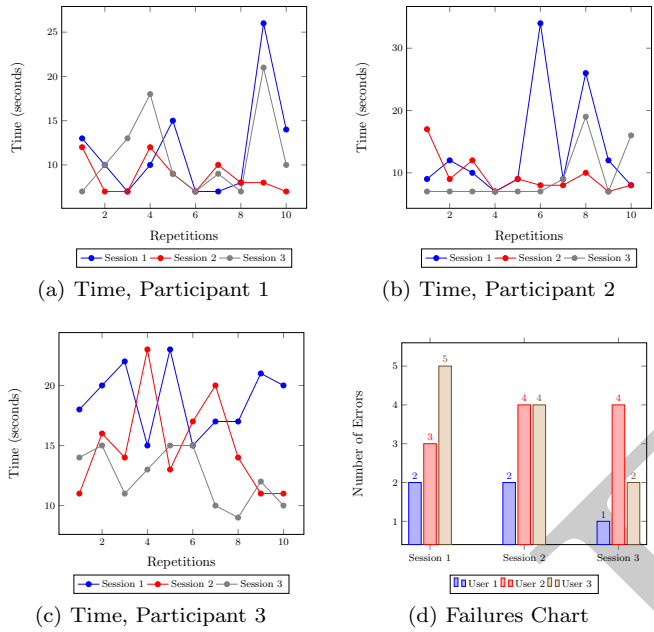

In Figures \ref{autismResults2nd}, \ref{hearingResults2nd}, \ref{physicalResults2nd} and \ref{visualResults2nd} the data obtained from this second experiment with the students are shown.


In this second iteration, it is important to compare the results with the first iteration in order to check the progress of the students and to see if the device-interaction model is working. The most important characteristic in this second phase is that the time taken and the number of errors have decreased with respect to the first iteration. The charts relating to the time performance (see Figures \ref{autismResults2nd}, \ref{hearingResults2nd}, \ref{physicalResults2nd} and \ref{visualResults2nd}) show the reduction in time. With regards to the decrease in the number of errors in this second iteration (see Figure \ref{autismResults2nd_errors}, \ref{hearingResults2nd_errors}, \ref{physicalResults2nd_errors}, \ref{visualResults2nd_errors}) it is deduced that the users have understood the purpose of the tasks and interaction with the system was no longer an obstacle to achieving the goals. It is necessary to say that none of the users had any previous experience with the interaction device and this explains why the results were very slow and irregular at the start of the first session in the previous iteration. This is not evident in the present iteration. The most outstanding values are the ones presented below:

\begin{itemize}
\item In the sessions with the users with severe hearing loss, the time and the number of errors decreased and the results were uniform. For this evaluation, pictures were added instead of text and the tutor’s instructions were not necessary.
\item It was observed that the students with physical disability seemed more indecisive when they interacted with the system. Furthermore, their movements were more limited and as a result, their performance time was slower than the rest of the participants. This showed that, although the distance between the elements has been reduced to make the interaction easier, it would have to be further  reduced, especially in the cases where the user is confined to a wheelchair.
\item In the evaluation with the students with autism, it can be concluded that they find it easier to interact with the system than the rest of the participants. The main reason is that these students do not have any limitation in their movements or any of a sensory aspect. The only drawback is that they may lose attention and interest more easily. This is why the methodology for the activities in these students is so different with regard to other cases. 
\end{itemize}

\begin{figure} [!h]
\centering
\begin{tikzpicture}[scale=0.7]
	\begin{axis}[
		xlabel=Repetitions,ylabel=Time (seconds), legend style={at={(0.5,-0.2)},
      anchor=north,legend columns=-1}, label style={font=\large}]

	\addplot[color=blue,mark=*] coordinates {
		(1,25)
		(2,28.16)
		(3,20.05)
		(4,17.25)
		(5,19.5)
		(6,20.16)
		(7,20.16)
		(8,20.5)
		(9,20.4)
		(10,26.58)
	};
	\addplot[color=red,mark=*] coordinates {
		(1,22.3)
		(2,17.08)
		(3,16.66)
		(4,20.25)
		(5,13.5)
		(6,17)
		(7,18)
		(8,13.75)
		(9,15)
		(10,13.33)
	};
	\addplot[color=gray,mark=*] coordinates {
		(1,17.5)
		(2,14.41)
		(3,13.3)
		(4,24.6)
		(5,13.33)
		(6,14.91)
		(7,15.75)
		(8,13.41)
		(9,13.08)
		(10,12.25)
	};
	\legend{Session 1, Session 2, Session 3}
	\end{axis}%
\end{tikzpicture}%
\caption{General Time Results $1^{st}$ iteration.}
\label{generalTimeResults1st}
\end{figure}
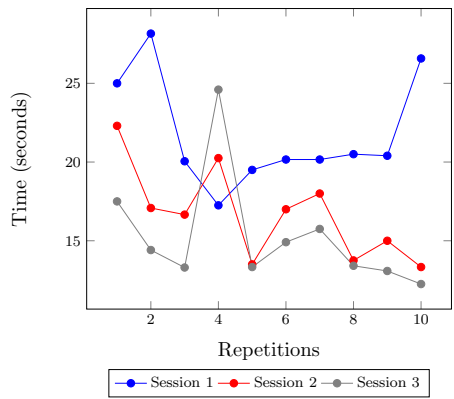

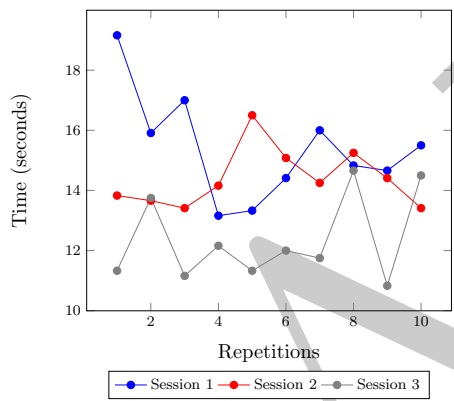
\begin{figure} [!h]
\centering
\begin{tikzpicture}[scale=0.7]
	\begin{axis}[
		xlabel=Repetitions,ylabel=Time (seconds), legend style={at={(0.5,-0.2)},
      anchor=north,legend columns=-1}, label style={font=\large}]

	\addplot[color=blue,mark=*] coordinates {
		(1,19.16)
		(2,15.91)
		(3,17)
		(4,13.16)
		(5,13.33)
		(6,14.41)
		(7,16)
		(8,14.83)
		(9,14.66)
		(10,15.5)
	};
	\addplot[color=red,mark=*] coordinates {
		(1,13.83)
		(2,13.66)
		(3,13.41)
		(4,14.16)
		(5,16.5)
		(6,15.08)
		(7,14.25)
		(8,15.25)
		(9,14.41)
		(10,13.41)
	};
	\addplot[color=gray,mark=*] coordinates {
		(1,11.33)
		(2,13.75)
		(3,11.16)
		(4,12.16)
		(5,11.33)
		(6,12)
		(7,11.75)
		(8,14.66)
		(9,10.83)
		(10,14.5)
	};
	\legend{Session 1, Session 2, Session 3}
	\end{axis}%
\end{tikzpicture}%
\caption{General Time Results $2^{nd}$ iteration.}
\label{generalTimeResults2nd}
\end{figure}

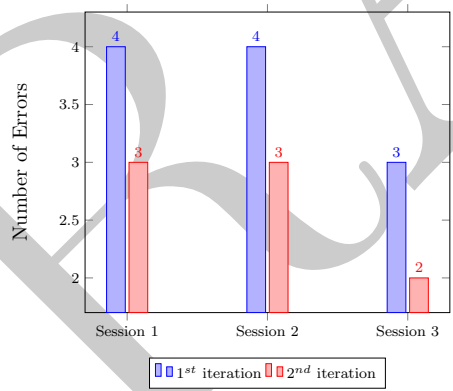
\begin{figure} [!h]
\centering
\begin{tikzpicture}[scale=0.7]
\begin{axis}[
    ybar,
    enlargelimits=0.15,
    legend style={at={(0.5,-0.15)},
      anchor=north,legend columns=-1},
    ylabel={Number of Errors},
    symbolic x coords={Session 1,Session 2,Session 3},
    xtick=data,
    nodes near coords,
    nodes near coords align={vertical},
    label style={font=\large}
    ]
\addplot coordinates {(Session 1,4) (Session 2,4) (Session 3,3)};
\addplot coordinates {(Session 1,3) (Session 2,3) (Session 3,2)};
\legend{$1^{st}$ iteration, $2^{nd}$ iteration}
\end{axis}
\end{tikzpicture}
\caption{General Errors Results}
\label{generalErrorResults}
\end{figure}

After explaining the individual results according to the characteristics of the participants, the data are grouped to have an overview of the users' evaluation. In order to do this, the mean times of the students with autism, visual impairment, physical disability and severe hearing loss were calculated and two graphs were obtained: a representative graph of the first iteration (see Figure \ref{generalTimeResults1st}) and another one for the second iteration (see Figure \ref{generalTimeResults2nd}). In addition, the average of the errors in both iterations was also calculated (see Figure \ref{generalErrorResults}).

Figure \ref{generalTimeResults1st} displays the results of the students' execution times, regardless of their characteristics, where it can be seen that the times are reduced in the consecutive sessions in which the last session is the one that presents the lowest values. Therefore, from a general point of view, the participants have got used to interacting with the system and have been able to complete the exercises in a shorter time.

On the other hand, if we look at the lines of the graph in Figure \ref{generalTimeResults2nd}, the same conclusions in the previous chart could be obtained since the execution time decreases as the sessions progress. However, the significant aspect of this graph lies in the y-axis because the scale is smaller than in the previous graph, which shows that the participants are doing the exercises in a shorter time. Although in the first iteration the use of Kinect together with a new mode of interaction could pose a greater obstacle, the students are eventually able to interact with the system and feel comfortable with the help of the device-interaction model. In fact, it adjusts to the needs of each individual to facilitate interaction with the different activities proposed. Moreover, the error chart (see Figure \ref{generalErrorResults}) shows that the numbers of failures are lower in the second iteration than in the first one, similar to the results where the data is separated by disabilities. 

\begin{table*} [!h]
\renewcommand{\arraystretch}{1.3}
\caption{Statistic Parameters about Time of all Participants (SD: Standard Deviation; CV: Coefficient of Variation).}
\label{table_stats}
\centering
{\fontsize{8}{9}\selectfont
\begin{tabular}{|l|c|c|c|c|c|}
\hline
{\bf \#Disability}  & {\bf User} & {\bf \#iterat.} & {\bf Average} & {\bf SD} & {\bf CV}  \\
\hline
\multirow{6}{*}{Autism} & \multirow{2}{*}{1} & 1 & 13,16 & 7,49 & 0,56 \\
 &  & 2 & 11,73 & 3,83 & 0,32 \\
 &  \multirow{2}{*}{2} & 1 & 25,7 & 18,51 & 0,72  \\
& & 2 & 15,26 & 6,64 & 0,43  \\
 &  \multirow{2}{*}{3} & 1 & 18,46 & 6,19 & 0,33 \\
&  & 2 & 14,7 & 2,79 & 0,19 \\
\hline
\multirow{6}{*}{Hearing Impairment} & \multirow{2}{*}{1} & 1 & 15,96 & 11,26 & 0,70 \\
 &  & 2 & 15,26 & 6,36 & 0,41 \\
 &  \multirow{2}{*}{2} & 1 & 18,63 & 9,86 & 0,52  \\
& & 2 & 16 & 5,84 & 0,36  \\
 &  \multirow{2}{*}{3} & 1 & 17,5 & 5,02 & 0,28 \\
&  & 2 & 14,7 & 4,75 & 0,19 \\
\hline
\multirow{6}{*}{Physical Impairment} & \multirow{2}{*}{1} & 1 & 13,96 & 10,68 & 0,76 \\
 &  & 2 & 11,76 & 5,69 & 0,48 \\
 &  \multirow{2}{*}{2} & 1 & 21,13 & 18,03 & 0,85  \\
& & 2 & 14,86 & 15,18 & 1  \\
 &  \multirow{2}{*}{3} & 1 & 18,5 & 14,82 & 0,80 \\
&  & 2 & 12,83 & 6,34 & 0,49 \\
\hline
\multirow{6}{*}{Visual Impairment} & \multirow{2}{*}{1} & 1 & 9,23 & 2,67 & 0,28 \\
 &  & 2 & 10,5 & 4,56 & 0,43 \\
 &  \multirow{2}{*}{2} & 1 & 19,5 & 5,44 & 0,27  \\
& & 2 & 10,8 & 6,16 & 0,57  \\
 &  \multirow{2}{*}{3} & 1 & 25,1 & 4,40 & 0,17 \\
&  & 2 & 15,4 & 4,04 & 0,26 \\
\hline
\end{tabular}
}
\end{table*}

Finally, Table \ref{table_stats} shows the average, standard deviation and coefficient of variation of the times obtained from the different sessions that have been carried out in the two iterations. The average on its own is not entirely relevant since there may be a divergence in the values, which would mean that some of the activity parameters such as the interactive mode or the student does not understand the instructions. These circumstances would cause the students to lose confidence and become ill at ease with  the activities and consequently bring about these drastic variations in time. Thus, the coefficient of variation was calculated and knowing the homogeneity of the values, the nearer the values to are zero, the more homogeneity there will be. This coefficient is closer to zero in the second iteration, hence we can assume the users understand the dynamics of the activity and they are able to carry it out with the model proposed. Moreover, this serves to support  the decision to change the room to do the activities because of the distractions in the previous location since the coefficient of variation is lower in the second iteration and the values are more lineal with a minority of high values compared to the first session.

\subsection{User Experience Questionnaire (UEQ)}

Finally, we carried  out a qualitative  assessment with  the five tutors  of the students who participated in the previous evaluation. The tutors had to fill the User Experience Questionnaire (UEQ) \cite{laugwitz2008construction} according to their experience with the system. The UEQ is an efficient method to collect the users’ opinions regarding their experience of a product \cite{laugwitz2008construction}. The main goal of the UEQ is to offer a fast way to measure the user experience in a product. Furthermore, the reliability and validity of the UEQ has been tested with 144 participants and an online survey with 722 participants, providing a quality assessment tool \cite{schrepp2014applying}. This questionnaire contains 26 items (see Figure \ref{fig:questionnaire_template}) that are divided in 6 scales presented below:

 \begin{itemize}
    \item [---] {\bf Attractiveness}: It identifies whether the users like or dislike the product.
    \item [---] {\bf Perspicuity}: It verifies if it is easy to learn how to use the product.
    \item [---] {\bf Efficiency}: It demonstrates if the users solve the tasks without much effort.
    \item [---] {\bf Dependability}: It validates whether the users feel comfortable with the interaction.
    \item [---] {\bf Stimulation}: It identifies if the users feel motivated while they use the product.
    \item [---] {\bf Novelty}: It demonstrates whether the product is innovative.
 \end{itemize}

\begin{figure} [!h]
    \centering

\QItem{ \Qq{Item} \Qtab{2.32cm}{annoying \Qrating{7} enjoyable}}

\QItem{ \Qq{Item} \Qtab{0.1cm}{not understandable \Qrating{7} understandable}}

\QItem{ \Qq{Item} \Qtab{2.49cm}{creative \Qrating{7} dull}}

\QItem{ \Qq{Item} \Qtab{1.84cm}{easy to learn \Qrating{7} difficult to learn}}

\QItem{ \Qq{Item} \Qtab{2.45cm}{valuable \Qrating{7} inferior}}

\QItem{ \Qq{Item} \Qtab{2.68cm}{boring \Qrating{7} exciting}}

\QItem{ \Qq{Item} \Qtab{1.63cm}{not interesting \Qrating{7} interesting}}

\QItem{ \Qq{Item} \Qtab{1.76cm}{unpredictable \Qrating{7} predictable}}

\QItem{ \Qq{Item} \Qtab{3.06cm}{fast \Qrating{7} slow}}

\QItem{ \Qq{Item} \Qtab{2.39cm}{inventive \Qrating{7} conventional}}

\QItem{ \Qq{Item} \Qtab{2.1cm}{obstructive \Qrating{7} supportive}}

\QItem{ \Qq{Item} \Qtab{2.92cm}{good \Qrating{7} bad}}

\QItem{ \Qq{Item} \Qtab{2cm}{complicated \Qrating{7} easy}}

\QItem{ \Qq{Item} \Qtab{2.4cm}{unlikable \Qrating{7} pleasing}}

\QItem{ \Qq{Item} \Qtab{2.87cm}{usual \Qrating{7} leading edge}}

\QItem{ \Qq{Item} \Qtab{2.14cm}{unpleasant \Qrating{7} pleasant}}

\QItem{ \Qq{Item} \Qtab{2.74cm}{secure \Qrating{7} not secure}}

\QItem{ \Qq{Item} \Qtab{2.17cm}{motivating \Qrating{7} demotivating}}

\QItem{ \Qq{Item} \Qtab{1.1cm}{meets expectations \Qrating{7} does not meet expectations}}

\QItem{ \Qq{Item} \Qtab{2.31cm}{inefficient \Qrating{7} efficient}}

\QItem{ \Qq{Item} \Qtab{2.97cm}{clear \Qrating{7} confusing}}

\QItem{ \Qq{Item} \Qtab{2.15cm}{impractical \Qrating{7} practical}}

\QItem{ \Qq{Item} \Qtab{2.35cm}{organized \Qrating{7} cluttered}}

\QItem{ \Qq{Item} \Qtab{2.35cm}{attractive \Qrating{7} unattractive}}

\QItem{ \Qq{Item} \Qtab{2.6cm}{friendly \Qrating{7} unfriendly}}

\QItem{ \Qq{Item} \Qtab{2cm}{conservative \Qrating{7} innovative}}

    \caption{Questionnaire (The values are assigned from 1 to 7).}
    \label{fig:questionnaire_template}
\end{figure}

Although a short version of the UEQ which contained only 8 items \cite{schrepp2017design}, was drawn up as some of the items were not relevant for some products, we decided to use the standard version since we thought that every item fitted with our system and was able to help improve the results for our study.

The main advantages of applying this method are:
\begin{itemize}
    \item [a)] It takes into account these three criteria regarding the user experience:
    \begin{itemize}
        \item  Their feelings about the interaction with the product by the standard ISO 9241-10 \cite{din19969241}.
        \item  The effectiveness or efficiency for ISO 9241-11 \cite{iso1998ergonomic}.
        \item  The user satisfaction related to hedonic quality \cite{iso1998ergonomic}.
    \end{itemize}
    \item [b)] It is simple and fast.
    \item [c)] A benchmark was created to improve accuracy.
    \item [d)] It has been translated to many languages, thereby facilitating its use and making the results more reliable.
\end{itemize}

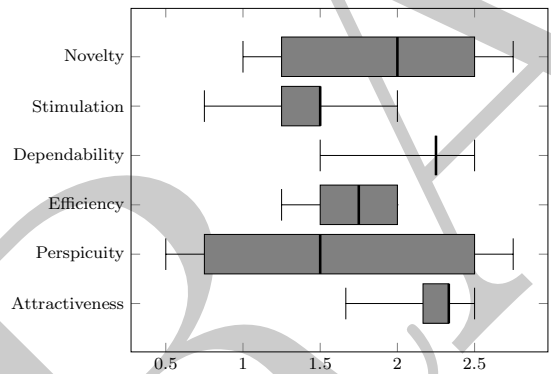
\begin{figure} [!h]
    \centering
    \caption{UEQ Scales Chart.}
    \label{fig:boxplot}
\begin{tikzpicture}[scale=0.8]
  \begin{axis}
    [
    ytick={1,2,3,4,5,6},
    yticklabels={Attractiveness, Perspicuity, Efficiency, Dependability, Stimulation, Novelty},
    ]
    \addplot+[
    boxplot prepared={
      median=2.333,
      upper quartile=2.333,
      lower quartile=2.166,
      upper whisker=2.5,
      lower whisker=1.666
    },
    ] coordinates {};
    \addplot+[
    boxplot prepared={
      median=1.5,
      upper quartile=2.5,
      lower quartile=0.75,
      upper whisker=2.75,
      lower whisker=0.5
    },
    ] coordinates {};
    \addplot+[
    boxplot prepared={
      median=1.75,
      upper quartile=2,
      lower quartile=1.5,
      upper whisker=2,
      lower whisker=1.25
    },
    ] coordinates {};
    \addplot+[
    boxplot prepared={
      median=2.25,
      upper quartile=2.25,
      lower quartile=2.25,
      upper whisker=2.5,
      lower whisker=1.5
    },
    ] coordinates {};
    \addplot+[
    boxplot prepared={
      median=1.5,
      upper quartile=1.5,
      lower quartile=1.25,
      upper whisker=2,
      lower whisker=0.75
    },
    ] coordinates {};
    \addplot+[
    boxplot prepared={
      median=2,
      upper quartile=2.5,
      lower quartile=1.25,
      upper whisker=2.75,
      lower whisker=1
    },
    ] coordinates {};
  \end{axis}
  
\end{tikzpicture}
\end{figure}

The five tutors filled the UEQ (see Table \ref{table_UEQ_answers}) anonymously according to the experience that they had had in the previous experiment with the students and from using the system themselves. Then, we analyzed the results with the data analysis tools that the authors provide in order to know how our system rates in terms of user experience. The results are displayed in Table \ref{table_UEQ_scales} and Figure \ref{fig:boxplot}.

\begin{table*} [!h]
\renewcommand{\arraystretch}{1.3}
\caption{User Experience Questionnaire Answers (Participants \#1 to \#5).}
\label{table_UEQ_answers}
\centering
{\fontsize{8}{9}\selectfont
\begin{tabular}{|cccccc|}
\hline
{\bf Item}  & {\bf\#P1} & {\bf\#P2} & {\bf\#P3} & {\bf\#P4} & {\bf\#P5} \\
\hline
1 & 6 & 7 & 6 & 5 & 6 \\
\hline
2 & 5 & 5 & 7 & 5 & 6 \\
\hline
3 & 3 & 5 & 3 & 3 & 1  \\
\hline
4 & 5 & 3 & 1 & 2 & 1  \\
\hline
5 & 2 & 4 & 3 & 2 & 2 \\
\hline
6 & 4 & 6 & 5 & 5 & 5 \\
\hline
7 & 6 & 5 & 6 & 5 & 6 \\
\hline
8 & 7 & 6 & 6 & 7 & 7 \\
\hline
9 & 3 & 5 & 4 & 2 & 3  \\
\hline
10 & 1 & 2 & 1 & 3 & 1  \\
\hline
11 & 6 & 7 & 6 & 5 & 6 \\
\hline
12 & 1 & 1 & 1 & 2 & 1 \\
\hline
13 & 5 & 6 & 6 & 6 & 7 \\
\hline
14 & 6 & 6 & 7 & 5 & 6 \\
\hline
15 & 7 & 6 & 6 & 5 & 7  \\
\hline
16 & 4 & 5 & 6 & 6 & 6  \\
\hline
17 & 2 & 2 & 3 & 1 & 1 \\
\hline
18 & 3 & 4 & 2 & 2 & 1 \\
\hline
19 & 2 & 2 & 3 & 2 & 2 \\
\hline
20 & 5 & 6 & 5 & 5 & 6 \\
\hline
21 & 3 & 2 & 2 & 6 & 1  \\
\hline
22 & 7 & 6 & 6 & 6 & 6  \\
\hline
23 & 2 & 1 & 2 & 1 & 1 \\
\hline
24 & 1 & 1 & 1 & 2 & 1 \\
\hline
25 & 1 & 2 & 2 & 2 & 2 \\
\hline
26 & 7 & 6 & 6 & 5 & 6 \\
\hline
\end{tabular}
}
\end{table*}

\begin{table*} [!h]
\renewcommand{\arraystretch}{1.3}
\caption{Scales for the UEQ}
\label{table_UEQ_scales}
\centering
{\fontsize{8}{9}\selectfont
\begin{tabular}{|ccc|}
\hline
{\bf Scale} & {\bf Mean} & {\bf Variance} \\
\hline
Attractiveness & 2.200 & 0.10 \\
\hline
Perspicuity & 1.600 & 1.02 \\
\hline
Efficiency & 1.700 & 0.11   \\
\hline
Dependability & 2.150 & 0.14 \\
\hline
Stimulation & 1.400 & 0.21  \\
\hline
Novelty & 1.900 & 0.58  \\
\hline
\end{tabular}
}
\end{table*}

The authors of the UEQ set a benchmark for this questionnaire because they thought that it would be useful, especially when the product is measured for the first time (as in this case), given the impossibility of making comparisons with previous evaluations \cite{schrepp2017construction}. Thus, we use it to have a reliable baseline for our analysis. According to the benchmark (see Table \ref{table_UEQ_benchmark}), the results that we obtained (see Table \ref{table_UEQ_scales} and Figure \ref{fig:boxplot}) from the questionnaire would be: Attractiveness Excellent, Perspicuity Good, Efficiency Good, Dependability Excellent, Stimulation Good, and Novelty Excellent.


\begin{table*} [!h]
\renewcommand{\arraystretch}{1.3}
\caption{Benchmark Borders for UEQ \cite{schrepp2017construction} (A: Attractiveness, P: Perspicuity, E: Efficiency, D: Dependability, S: Stimulation, N: Novelty).}
\label{table_UEQ_benchmark}
\centering
{\fontsize{9}{10}\selectfont
\begin{tabular}{lcccccc}
\hline
{\bf Category} & {\bf A} & {\bf P} & {\bf E} & {\bf D} & {\bf S} & {\bf N} \\
\hline
Excellent & $\geq$ 1.75 & $\geq$ 1.78 & $\geq$ 1.9 & $\geq$ 1.65 & $\geq$ 1.55 & $\geq$ 1.4 \\
\hline
\multirow{2}{*}{Good} & $\geq$ 1.52 & $\geq$ 1.47 & $\geq$ 1.56 & $\geq$ 1.48 & $\geq$ 1.31 & $\geq$ 1.05 \\
& $<$ 1.75 & $<$ 1.78 & $<$ 1.9 & $<$ 1.65 & $<$ 1.55 & $<$ 1.4 \\
\multirow{2}{*}{Above average} & $\geq$ 1.17 & $\geq$ 0.98 & $\geq$ 1.08 & $\geq$ 1.14 & $\geq$ 0.99 & $\geq$ 0.71   \\
& $<$ 1.52 & $<$ 1.47 & $<$ 1.56 & $<$ 1.48 & $<$ 1.31 & $<$ 1.05 \\
\hline
\multirow{2}{*}{Below average} & $\geq$ 0.7 & $\geq$ 0.54 & $\geq$ 0.64 & $\geq$ 0.78 & $\geq$ 0.5 & $\geq$ 0.3 \\
& $<$ 1.17 & $<$ 0.98 & $<$ 1.08 & $<$ 1.14 & $<$ 0.99 & $<$ 0.71 \\
\hline
Bad & $<$ 0.7 & $<$ 0.54 & $<$ 0.64 & $<$ 0.78 & $<$ 0.5 & $<$ 0.3  \\
\hline
\end{tabular}
}
\end{table*}

As can be seen, attractiveness, dependability and novelty obtained the best results in the test. However, despite perspicuity, efficiency and stimulation getting lower results, these were still good enough. We deduce that the participants rated novelty really high because of the use of Kinect to interact through gestures or motion detection with the application, especially as there is no other similar activity in the center. Regarding the score for attractiveness, the participants liked the application, thus it seems that the evaluation with experts was useful as it was following this that we changed some aspects in the interface and design. For this study the most important factor was dependability since this is related to interaction and our main goal in this study is to demonstrate that the device-interaction model which we have designed is useful and practical. Therefore, obtaining such a high rating in that scale is an excellent indication since it is clear the users thought that the system was suitably interactive. On the other hand, perspicuity has a lower score because the students are not used to interacting with Kinect and had to do a training process before the experiments to get accustomed. The lowest score corresponded to the stimulation scale even though we designed the activities as games in order to engage the students and to ensure they were not bored completing the tasks. We should check the design of the activities in order to motivate the students and perhaps add more activities to extend the variety.

\section{Conclusions and Future Work} \label{conclusions}

In this paper, the results for a device-interaction model were shown in which  the goal is to adapt the device-interaction model features to the user needs. In this particular case, Microsoft Kinect v2 was used, where the optimization features were the different components which are integrated: the RGB camera, the depth sensor and the array of microphones or the skeletal tracking.

This research, which also includes the device-interaction model, has user model based features and some adaptation rules which contain important information to enable the user to adapt to the system. This model, which is based on feature-value, improves the understanding and interpretation of the system since it has been integrated easily with the adaptation rules. This environment has an activity system that adapts the selected activity depending on the information stored in the user model and the device-interaction model.

Finally, the system was evaluated. First, an evaluation by experts and teachers was carried out, followed by another evaluation with 12 participants, which was segmented according to the different types of disabilities. In the second evaluation, it was noted that the performance time and the number of errors decreased which shows that every user, despite their disabilities, had managed to complete the proposed activities without any problem. Lastly, a qualitative evaluation using the User Experience Questionnaire was carried out in which the tutors participated in order to determine their experience with the application. This method evaluates the system according to six scales: Attractiveness, perspicuity, efficiency, dependability, stimulation and novelty. In particular the results showed that the users liked the application (attractiveness), that they did not have many problems with the interaction in spite of using Kinect (dependability) and they thought that the system was original. Though the rest of the categories obtained lower scores they were also  satisfactory. 

This research will enable the authors to develop future work in order to add new functionalities to the system and offer other alternatives, such as:
\begin{itemize}
\item [(a)] Developing the system to connect several devices to interact with the platform, such as Leap Motion, Intel RealSense, haptic devices and/or eye tracker.
\item [(b)] Creating a module which entails suggestions regarding devices to use according to the user profile. 
\item [(c)] Extending the field to include the cognitive aspect because the main goal of this work is focused on improving the interaction in users with physical or sensory disability.
\item [(d)] Creating a general module for a system that is able to convert an input signal from a device into an action. For example, in a home automation environment, the signal received by the device would be useful to control different home elements.
\end{itemize}

\begin{acknowledgements}
This work was supported by the EU ERDF and the Spanish Ministry of Economy and Competitiveness (MINECO) under Project TIN2017-83964-R. We would also like to thank Dr. Ana Garcia Serrano because of her collaboration in this study. We also want to thank to ARASAAC (http://arasaac.org) for allowing us to use Sergio Palao’s pictograms which belongs to the Aragon government.
\end{acknowledgements}

\bibliographystyle{spmpsci}      
\bibliography{references}   

\end{document}